\documentclass{pasj00}
\Received{2012 January 31}
\Accepted{2012 March 26}
\SetRunningHead{Yamada et al.}{Simulations of the SZE for ALMA}

\begin{document}
\title{Imaging Simulations of the Sunyaev-Zel'dovich Effect 
  for ALMA} 
\author{Kenkichi \textsc{Yamada}\altaffilmark{1},
Tetsu \textsc{Kitayama}\altaffilmark{1},
Shigehisa \textsc{Takakuwa}\altaffilmark{2},
Daisuke \textsc{Iono}\altaffilmark{3}, 
Takahiro \textsc{Tsutsumi}\altaffilmark{4}, \\
Kotaro \textsc{Kohno}\altaffilmark{5,6}, 
Motokazu \textsc{Takizawa}\altaffilmark{7},
Kohji \textsc{Yoshikawa}\altaffilmark{8},
Takuya \textsc{Akahori}\altaffilmark{9},
Eiichiro \textsc{Komatsu}\altaffilmark{10},\\
Yasushi \textsc{Suto}\altaffilmark{6,11,12},
Hiroshi \textsc{Matsuo}\altaffilmark{13}, 
and Ryohei \textsc{Kawabe}\altaffilmark{3,13,14}}

\altaffiltext{1}{Department of Physics, Toho University,  
Funabashi, Chiba 274-8510, Japan}
\altaffiltext{2}{Academia Sinica Institute of Astronomy and Astrophysics, P.O. Box 23-141, Taipei 10617, Taiwan}
\altaffiltext{3}{Nobeyama Radio Observatory, Minamimaki, Minamisaku, Nagano 384-1305, Japan}
\altaffiltext{4}{National Radio Astronomy Observatory, Socorro, NM 87801-0387, USA}
\altaffiltext{5}{Institute of Astronomy, University of Tokyo, 2-21-1 Osawa, Mitaka, Tokyo 181-0015, Japan}
\altaffiltext{6}{Research Center for the Early Universe, School of Science, 
The University of Tokyo, Tokyo 113-0033, Japan}
\altaffiltext{7}{Department of Physics, Yamagata University, Kojirakawa-machi 1-4-12, Yamagata 990-8560, Japan}
\altaffiltext{8}{Center for Computational Sciences, University of Tsukuba, 1-1-1, Tennodai, Ibaraki 305-8577, Japan}
\altaffiltext{9}{ Korea Astronomy and Space Science Institute,
Daedeokdaero 776, Yuseong, Daejeon 305-348, Korea}
\altaffiltext{10}{Texas Cosmology Center and Department of Astronomy,
The University of Texas at Austin, Austin, TX 78712, USA}
\altaffiltext{11}{Department of Physics, The University of Tokyo,
Tokyo 113-0033, Japan}
\altaffiltext{12}{Department of Astrophysical Sciences, Princeton
University, Princeton, NJ 08544, USA}
\altaffiltext{13}{National Astronomical Observatory of Japan, 2-21-1 Osawa, Mitaka, Tokyo 181-8588, Japan}
\altaffiltext{14}{Joint ALMA Observatory, Alonso de Cordova 3107 OFC 129, Vitacura, Chile}


\maketitle

\begin{abstract}
We present imaging simulations of the Sunyaev-Zel'dovich effect of
galaxy clusters for the Atacama Large Millimeter/submillimeter Array
(ALMA) including the Atacama Compact Array (ACA).  In its most compact
configuration at 90GHz, ALMA will resolve the intracluster medium with
an effective angular resolution of 5 arcsec. It will provide a unique
probe of shock fronts and relativistic electrons produced during cluster
mergers at high redshifts, that are hard to spatially resolve by current
and near-future X-ray detectors.  Quality of image reconstruction is
poor with the 12m array alone but improved significantly by adding ACA;
expected sensitivity of the 12m array based on the thermal noise is not
valid for the Sunyaev-Zel'dovich effect mapping unless accompanied by an
ACA observation of at least equal duration. The observations above 100
GHz will become excessively time-consuming owing to the narrower
beam size and the higher system temperature. On the other hand,
significant improvement of the observing efficiency is expected once
Band 1 is implemented in the future.

\end{abstract}
\section{Introduction}

The Sunyaev-Zel'dovich effect (SZE, \cite{Sunyaev72}), inverse Compton
scattering of the cosmic microwave background off hot electrons,
provides a unique probe of the intracluster medium (for reviews see
\cite{Rephaeli95, Birkinshaw99, Carlstrom02}). For the gas with given
electron density $n_{\rm e}$ and temperature $T_{\rm e}$ at redshift
$z$, surface brightness of the thermal SZE is proportional to the
line-of-sight integral of electron density times temperature, $\int
n_{\rm e} T_{\rm e} dl$, whereas that of the X-ray thermal
bremsstrahlung emission is proportional to $(1+z)^{-4}\int n_{\rm e}^2
T_{\rm e}^{1/2} dl$.  The relative significance of the SZE over the
X-ray emission thus increases with redshift and electron
temperature. This makes the SZE a powerful tool in detecting the shocks
and the hot gas associated with violent cluster mergers
(\cite{Kitayama04, Ota08, Korngut11}), the frequency of which is
expected to increase at high redshifts when the growth of cosmic
structures was faster than today.

As of today, however, spatial resolutions of the majority of the SZE
images are at arcminute scales with a limited number of exceptions
(Komatsu et al. 1999, 2001; \cite{Pointecouteau01,Mason10,Zemcov10,
Massardi10, Korngut11}). This is largely due to the low surface
brightness of the SZE and a small number of radio telescopes with
sub-arcminute resolution in the range of wavelengths relevant to the SZE
measurements.  Controlling systematics of such large telescopes to the
sensitivity level required for the SZE imaging is also challenging.
Interferometers are therefore a promising and complementary tool for
high sensitivity SZE imaging observations (e.g., \cite{Jones93,
Carlstrom96, AMI06, Muchovej07, Wu09}).  Since it measures synchronized
and correlated signals among different telescopes, an interferometer has
a much better control of systematic noise, e.g., from the atmosphere.
The obtained data also span a wide range of spatial scales, making it
possible to separate point-like sources from the SZE. On the other hand,
it often has a limited field-of-view of $0.5' \sim 2'$ in the
millimeter/submillimeter bands and requires a large number of pointings
(or mosaics) to uncover the bulk of the cluster emission that extends
over several arcminutes. The sensitivity also varies with spatial scales
depending on observing conditions such as array configuration.  Addition
of single-dish data further improves the sensitivity for extended
sources (e.g., \cite{Vogel84,Stanimirovic99,Takakuwa03,Kurono09}). It is
hence not always easy to know a priori to what extent a specific
interferometer can be used to image the SZE within a realistic observing
time.

In this paper, we investigate the feasibility of SZE imaging
observations by the Atacama Large Millimeter/submillimeter Array
(ALMA).\footnote{http://www.almascience.org/} ALMA will consist of at
least 66 antennas, designed to operate at photon frequencies between
$30$ GHz and $950$ GHz.  A major array of 12-meter antennas will be
combined with the Atacama Compact Array (ACA, \cite{Iguchi09}) made up
of 7-meter antennas and 12-meter single-dish antennas to improve the
spatial frequency coverage and the total power measurement. While the
inclusion of ACA should greatly enhance the capability of ALMA in
observing extended sources, the practical feasibility and how to
optimize it with such heterogeneous arrays in real observations are not
straightforward and certainly depend on the nature of the target.  We
thus perform detailed imaging simulations of the SZE observations with
ALMA including ACA using their latest configurations (see \cite{Pety01,
Helfer02, Pfrommer05, Takakuwa08} for earlier imaging simulations of
different targets).

We pay particular attention to resolving the shock structures of distant
merging clusters that are relatively compact and hence suit the
field-of-view of ALMA. We take the so-called bullet cluster, 1E 0657-558
at $z=0.296$, as a representative example and perform mock observations
using snap shots of state-of-the-art numerical simulations for this
cluster \citep{Takizawa05, Akahori12}.  1E 0657-558 is well known for
its prominent bow shock observed by Chandra \citep{Markevitch02} with
the inferred Mach number of $3.0 \pm 0.4$ and the shock velocity of
$\sim 4700$ km s$^{-1}$ \citep{Markevitch07}. It has also been observed
frequently via the SZE (e.g., \cite{Andreani99, Halverson09, Plagge10,
Zemcov10,Malu10}). The temperature or the energy distribution of the
shocked electrons, however, is still poorly constrained even for this
well-observed cluster because of the lack of sensitivity of current
X-ray and SZE detectors. We will show that ALMA is a powerful instrument
capable of resolving a merger shock structure of galaxy clusters at high
redshifts and improving our understanding of cluster evolution.

Throughout the paper, we assume a standard set of cosmological
parameters: $\Omega_m=0.27$, $\Omega_\Lambda=0.73$, and $h=0.70$
\citep{Komatsu11}.  In this cosmology, an angular size of 1$''$
corresponds to a physical size of 4.4 kpc and 8.1 kpc at $z=0.3$ and
$z=1$, respectively.

\section{Imaging Simulations}

The imaging simulations are performed using the MIRIAD software
(\cite{Sault95}), version 4.2.3. Following the currently planned
specifications of ALMA, we consider an array made up of fifty 12-meter
antennas (hereafter 12m$\times$50) and ACA made up of twelve 7-meter
antennas (hereafter 7m$\times$12) and four 12-meter single-dish antennas
(hereafter SD$\times$4). A summary of telescope specifications are
given in Table \ref{tab-telesc}.

\begin{figure}[t]
  \begin{center}
    \FigureFile(80mm,50mm){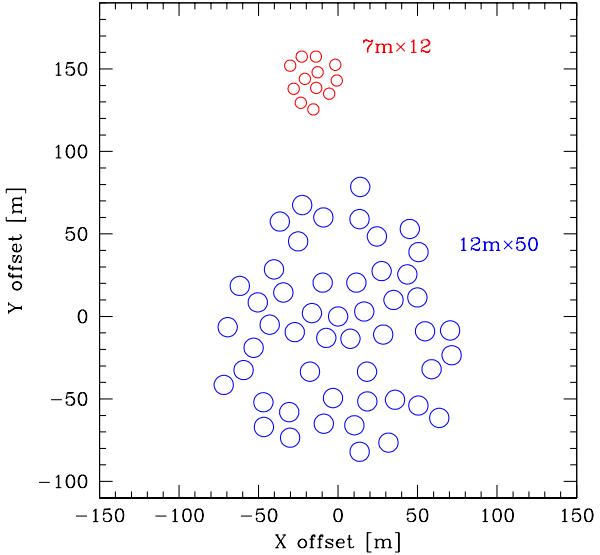}
  \end{center}
  \caption{The compact array configuration of 12m$\times$50 and
7m$\times$12 used in our mock observations. The projected positions on
the UTM coordinate are shown with the symbol sizes representing the
diameters of the telescopes. }  \label{fig-antenna}
\end{figure}

To maximize the sensitivity for the extended signals, the most compact
array configurations, shown in Figure \ref{fig-antenna}, are adopted for
12m$\times$50 and 7m$\times$12.\footnote{The antenna configuration files
are identical to alma.out01.cfg and aca.i.cfg used in CASA 3.3.}
The simulations are performed at 90GHz, which is the lowest frequency
(and hence has the largest field-of-views and the lowest system
temperature) in the initially planned specifications of ALMA. This
frequency is also expected to be an optimal choice for the SZE
observation owing to the minimal contamination by
synchrotron and dust emissions. The prospects for the observations in
the other bands will be discussed in \S \ref{sec-discuss}.

\begin{table*}[t]
\caption{Summary of telescope parameters.}
\begin{center}
\begin{tabular}{ll}
\hline\hline
Number of antennas & 12m$\times$50, 7m$\times$12, SD$\times$4 \\ 
Configuration & Most compact \\
Frequency & 90 GHz (Band 3) \\
Bandwidth   & 7.5 GHz per polarization\\
Intermediate frequency & 6 GHz \\ 
System temperature & 73.5 K \\
Aperture efficiency & 0.71 \\
Correlator efficiency & 0.88 \\
Primary beam FWHM & 69$''$ (12m$\times$50), ~118$''$ (7m$\times$12) \\
Phase error (rms) & $20^{\circ}$ \\
Gain error (rms) & $0.1\%$ \\
Pointing error (rms) & $0.6''$ 
\\\hline 
\end{tabular} 
\end{center}
\label{tab-telesc} 
\end{table*}

\subsection{Input Models}
\label{sec-input}

We use three input models whose observational set-ups are summarized in
Table \ref{tab-models}.  For each of the model images, we adopt the
pixel size of 1$''$, to satisfy the Nyquist condition for the smallest
spatial scale, $\sim 3''$ full width at half maximum (FWHM), probed by
the longest baseline of ALMA at 90GHz in the most compact
configuration. Wherever necessary, linear interpolation is performed to
assign the intensity to each pixel. The model images are created to
cover the entire observing area.

\begin{table*}[t]
\caption{Summary of model parameters. 
}
\begin{center}
\begin{tabular}{llll}
\hline\hline
Parameter &Model A&  Model B& Model C  \\
\hline
RA (J2000) & $00:00:00$ & $6:58:27.4$ & $6:58:17.3$\\
DEC (J2000) & $-23:00:00$ & $-55:56:50$ & $-55:56:30$\\
Hour Angle & $-1$ hr $\sim$ 1 hr   & $-5$ hr $\sim$ 5 hr & $-5$ hr $\sim$ 5 hr\\
Mapping area &&& \\
 \hspace*{4mm}12m$\times$50, 7m$\times$12 & $120''\times 120''$ & $240''\times 240''$ & $120''\times 120''$\\
 \hspace*{4mm}SD$\times$4 & $190''\times 190''$ & $310''\times 310''$ & $190''\times 190''$\\
Pixel size &&& \\
 \hspace*{4mm}12m$\times$50, 7m$\times$12 & $1''$
& $1''$& $1''$\\
 \hspace*{4mm}SD$\times$4 & $10''$& $10''$ & $10''$\\
Number of mosaics/pointings &   &  &  \\
 \hspace*{4mm} 12m$\times$50  &  19 & 67 & 19 \\
  \hspace*{4mm}  7m$\times$12 & 7 & 23 & 7 \\
  \hspace*{4mm}  SD$\times$4 & 400 & 1024 & 400 \\
Total integration time &   &  & \\
\hspace*{4mm} 12m$\times$50&  
2 hr & 10 hr & 10 hr \\
\hspace*{4mm} 7m$\times$12, SD$\times$4
&  2 hr & 40 hr & 40 hr\\
Noise added &   &  &  \\
 \hspace*{4mm}12m$\times$50, 7m$\times$12 & thermal  & thermal, phase & thermal, phase \\
       &           & gain, pointing & gain, pointing \\
\hspace*{4mm}SD$\times$4  & thermal  & thermal & thermal \\
Synthesized beam (major, minor, p.a.)\footnotemark[$*$] &   &  &  \\
\hspace*{4mm} 12m$\times$50  &  $4.0'', ~3.8'', ~-7.3^{\circ}$
& $4.9'',~4.2'',~-88^{\circ} $ & $4.8'',~4.2'',~-88^{\circ} $ \\
  \hspace*{4mm}  7m$\times$12 & $17'',~16'', ~5.1^{\circ}$ & $19'',~18'',~83^{\circ} $ & $19'',~18'',~84^{\circ} $ \\
  \hspace*{4mm} 12m$\times$50+7m$\times$12 & $4.1'', ~3.8'', ~-7.3^{\circ}$ & 
$5.2'',~4.6'',~-88^{\circ}$  & $5.2'',~4.5'',~-88^{\circ}$  \\
  \hspace*{4mm} SD$\times$4 & $69'', ~69'', ~0^{\circ}$ & 
$69'', ~69'', ~0^{\circ}$ & $69'', ~69'', ~0^{\circ}$ \\
Thermal noise ($\mu$Jy/arcsec$^2$, $\mu$Jy/beam)\footnotemark[$\dagger$] & & & \\
 \hspace*{4mm} 12m$\times$50  & 1.1, ~19  & 0.62, ~15& 0.35, ~8.0 \\
  \hspace*{4mm}  7m$\times$12 & 0.48, ~150& 0.15, ~60& 0.099, ~38 \\
  \hspace*{4mm}  12m$\times$50+7m$\times$12 & 1.0, ~18& 0.52, ~14& 0.30, ~7.8 \\
  \hspace*{4mm}  SD$\times$4 & 0.64, ~3400&  0.23, ~1200&  0.14, ~770 \\
\hline 
\multicolumn{4}{@{}l@{}}{\hbox to 0pt{\parbox{180mm}{\footnotesize
       \par\noindent
       \footnotemark[$*$] Three numbers for each model denote the major axis FWHM, the minor axis FWHM, and the position angle, respectively. 
       \par\noindent
       \footnotemark[$\dagger$] Two numbers for each  model denote 
the rms values in $\mu$Jy/arcsec$^2$ and $\mu$Jy/beam, respectively.
     }\hss}}
\end{tabular} 
\end{center}
\label{tab-models} 
\end{table*}

\subsubsection{Model A: Gaussian}

The first model is a simple two dimensional Gaussian, placed at zenith
$(\mbox{RA},\mbox{DEC}) =(00:00:00,-23:00:00)$, with FWHM ranging from
$5''$ to $60''$.  The peak intensity is fixed at $-100
\mu$Jy/arcsec$^2$, corresponding to the Compton y-parameter of $5 \times
10^{-3}$ at 90GHz.  We use this model simply to examine how the
feasibility depends on the source size when the noises are negligible.

\subsubsection{Model B: Bullet cluster}

The second model is a latest result of 3D hydrodynamic simulations
of a cluster merger by \citet{Akahori12} that are designed to reproduce
the observed properties of the bullet cluster, 1E 0657-558. They carried
out a set of N-body and SPH simulations of a collision of two galaxy
clusters with virial masses of $1.5\times 10^{15}$ and $2.5\times
10^{14}$ M$_\odot$ and the initial relative velocity of 3000 km s$^{-1}$, using
a code developed by \citet{Akahori10}. We use the result after 1.12 Gyr
from the initial condition for their two temperature model and compute
the intensity map of the SZE, including relativistic corrections
\citep{Itoh04,Nozawa05}. To improve the agreement with the existing
data, we renormalize the electron density so that the peak y-parameter
($\propto n_{\rm e} T_{\rm e}$) matches the central value of $3.31 \times 10^{-4}$
inferred by APEX-SZ \citep{Halverson09}. This is realized by reducing
the electron density in each mesh by $26\%$.\footnote{Reducing only the
electron density yields the most conservative estimate of the predicted
SZE intensity for a given y-parameter, because the relativistic
correction leads to the larger reduction in the predicted intensity for
the higher electron temperature at 90GHz.}  To cover both the SZ
emission peak and the shock front, the image center is shifted by 
$-15''$ (66 kpc) and $+20''$ (88 kpc) along the right ascension and
declination, respectively, from that plotted in Figs 1--6 of
\citet{Akahori12} and placed at $(\mbox{RA},\mbox{DEC})
=(6:58:27.4,-55:56:50)$. As the spatial resolution of the simulation is
moderate near the shock front (26 kpc or 6.0$''$), we use this model
mainly to examine the feasibility of observing the global structure of a
real cluster.

\begin{table*}[t]
\caption{Positions and fluxes of model point sources. }
\label{tab:obs}
\begin{center}
\begin{tabular}{rrr|rrr}
\hline\hline
&Input&  && Reconstructed &  \\
$\Delta$R.A.[arcsec] & $\Delta$Dec [arcsec] & Flux [mJy] 
 & $\Delta$R.A.[arcsec] & $\Delta$Dec [arcsec] & Flux [mJy] 
\\ \hline
83  & $-12$ & 1.0   & 83& $-12$& 0.93\\
89 & $-35$ & 0.74   &89&$-35$& 0.69 \\
$-108$ & 147 & 0.47  &$-108$ &147&0.42\footnotemark[$*$]\\
144 & 86 & 0.45  &144&86& 0.43\footnotemark[$*$]\\
117 & $-111$ & 0.43  &117&$-111$& 0.37\\
$-68$ & $-111$ & 0.37  &$-68$&$-111$& 0.31\\
78 & 113 & 0.16  &79&113& 0.14\\
$-95$ & $-94$ & 0.13 &$-95$&$-94$ & 0.14 \\
$-25$ & 97 & 0.079 &$-25$&97& 0.092 \\
$-30$  & 9 &  0.035 &&&undetected \\
54  & $-68$ & 0.032 &&&undetected \\
1 & 44  & 0.028  &&&undetected\\
\hline
\multicolumn{4}{@{}l@{}}{\hbox to 0pt{\parbox{180mm}{\footnotesize
       \par\noindent
       \footnotemark[$*$] Sources lying outside the $240''\times 240''$ region mapped by 12m$\times$50. 
     }\hss}}
\end{tabular} 
\end{center}
\label{tab-sources} 
\end{table*}

To further examine the feasibility of separating radio sources from the
SZE, we add to Model B point sources in the field of 1E 0657-558
reported in the literature \citep{Liang00, Wilson08,
Malu10}.  The source flux at 90GHz is estimated by the following
procedure; i) wherever available, we extrapolate the observed fluxes at
the two nearest frequencies to 90GHz assuming a single power-law, ii) if
the observed flux is available only at one frequency, we extrapolate it
assuming a single power-law with a spectral index of $-0.5$, iii) to be
conservative against uncertainties in the extrapolation, we multiply the
extrapolated fluxes by a factor of 2. The resulting source properties
are listed in Table \ref{tab-sources}.

\subsubsection{Model C: Shock front}

The third model is based on a higher spatial resolution Eulerian mesh
simulations of a moving substructure within a main cluster by
\citet{Takizawa05}. The model used in this paper is nearly the same as
"the radial infall model" in Takizawa (2005) except that the gas density
is reduced by a factor of 10 so that the y-parameter around the shock
front matches that of Model B.  We compute the SZE map from a snapshot
at the elapsed time of 0.78 Gyr from the start of the simulation, which
roughly reproduces the observed morphology of 1E 0657-558, and place it
at $(\mbox{RA},\mbox{DEC}) =(6:58:17.3,-55:56:30)$.  This model predicts
the y-parameter gap across the shock front of $1.5 \times 10^{-4}$,
about half the central y-parameter of this cluster. As this simulation
has a higher spatial resolution (2 kpc or 0.5$''$) and a smaller box
size (0.8 Mpc or 180$''$) than \citet{Akahori12}, we use Model C to
examine the feasibility of resolving a merger shock in a cluster. As our
mock observations are performed on a slightly larger area than the box
size of Takizawa (2005), we pad mirror images to the original simulation
output to cover the entire observing area.

\subsection{Mock visibilities}
\label{sec-mock}

Mock interferometric observations are performed over the sky area of
$120''\times 120''$ (Models A and C) or $240''\times 240''$ (Model B).
Linear mosaicing (e.g., \cite{Sault96}) is performed with pointing
centers lying on triangular grids separated by half the primary beam
FWHM given in Table \ref{tab-telesc}.  Each mosaicing center is observed
repeatedly for 30 sec on-source and 5 sec toward an off-source
calibrator with a sampling interval of 5 sec. The numbers of mosaics,
the hour angles, and the total integration times are listed in Table
\ref{tab-models}. The target elevation is above 20$\degree$ throughout
the hour angles considered in this paper.  As shown in Appendix
\ref{sec-append}, effective integration time toward each sky point apart
from the map edge is nearly uniform and given by 
\begin{equation}
t_{\rm eff} 
\simeq 2.6 \frac{t_{\rm int}}{N_{\rm mos}}, 
\end{equation}
where $t_{\rm int}$ is the total integration time, $N_{\rm mos}$ is the
number of mosaics, and the numerical factor accounts for the overlap of
mosaics for the grid orientation mentioned above.

The visibility data are then created separately for 12m$\times$50 and
7m$\times$12 by performing the 2D Fourier transformation to the $u-v$
plane using the MIRIAD task {\it uvgen} and {\it uvmodel}. The
visibility loss due to shadowing by adjacent telescopes is also taken
into account. The thermal noise is computed in all models using the same
parameters as the ALMA sensitivity calculator
\footnote{http://almascience.nao.ac.jp/call-for-proposals/sensitivity-calculator}
as of March 2012; the system temperature, the aperture efficiency, the
correlator efficiency, the bandwidth, and the intermediate frequency are
listed in Table \ref{tab-telesc}. The system temperature is computed by
adopting the source declination of Model B ($-55:56:50$) and
precipitable water vapour of 2.748 mm at 90GHz.

For Models B and C, antenna phase and gain noises with rms values of 20
degrees and 0.1$\%$, respectively, are also added over an interval of 5
minutes using the MIRIAD task {\it gperror}. Relative pointing errors
are also added to these models in an approximate manner by convolving
the input images by a Gaussian with $0.6''$ rms before creating the
visibilities.  This would be sufficient for our present purpose because
the pointing errors practically affect our results only via
identifications of point sources and the $u-v$ coverage of ALMA is very
good as shown below.

Figure \ref{fig-uv} illustrates the $u-v$ coverage of our mock
observations.  The spatial frequencies that correspond to the baseline
lengths from 4 to 48 k$\lambda$ and 2 to 10 k$\lambda$ are covered
nearly uniformly by 12m$\times$50 and 7m$\times$12, respectively, where
$\lambda=3.3$ mm at 90GHz.  Properties of the resulting synthesized
beams are given in Table \ref{tab-models}. The effective area of the
synthesized beam is given by
\begin{equation}
A_{\rm beam} = \frac{2\pi \theta_{\rm beam}^2}{8 \ln2}  
= 
28 ~{\rm arcsec}^2 \left(\frac{\theta_{\rm beam}}{5''}\right)^2,
\end{equation}
where $\theta_{\rm beam}$ is the geometric mean of the major axis FWHM
and the minor axis FWHM.  

\begin{figure*}[t]
  \begin{center}
   \FigureFile(70mm,50mm){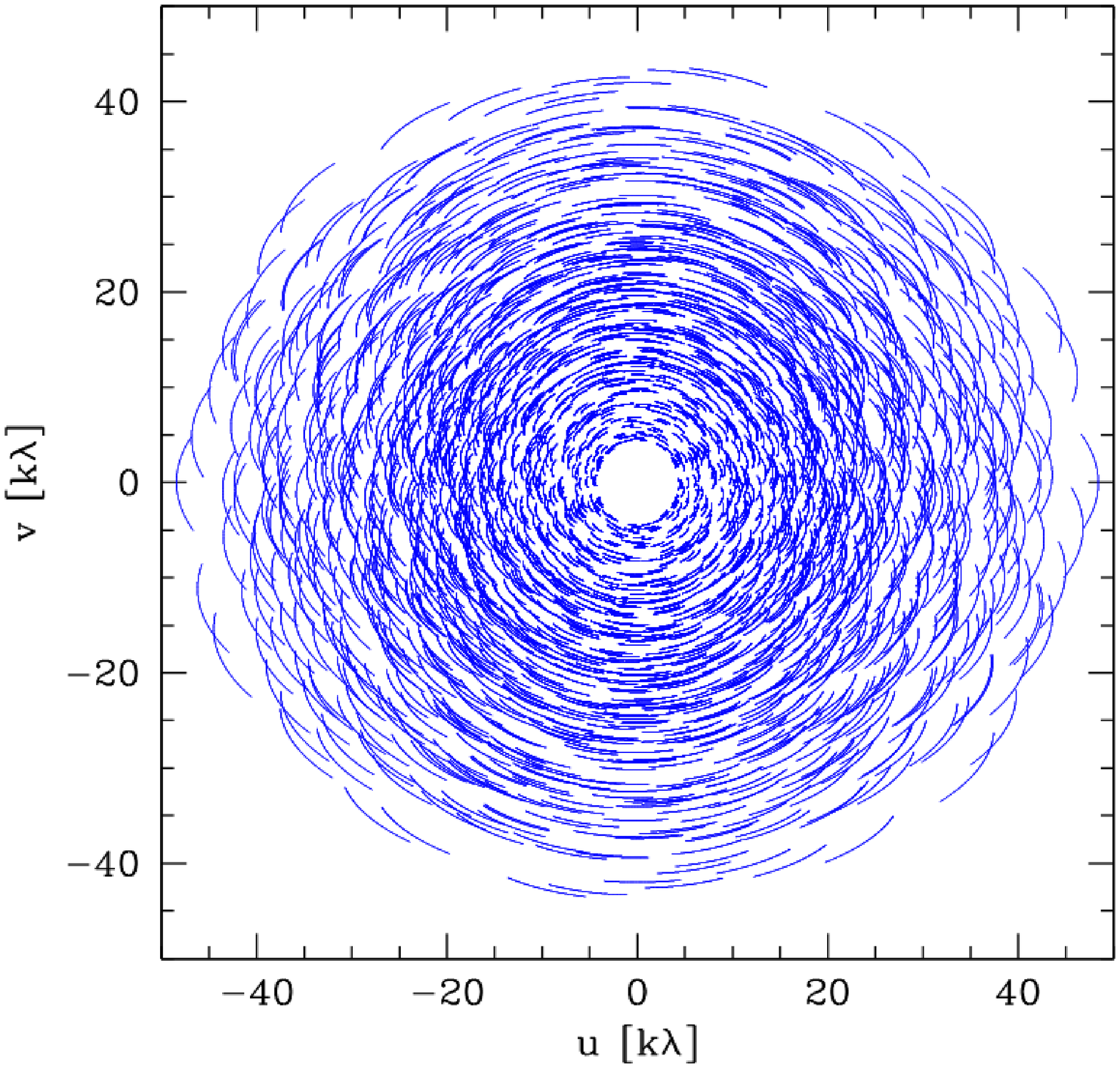}
   \FigureFile(70mm,50mm){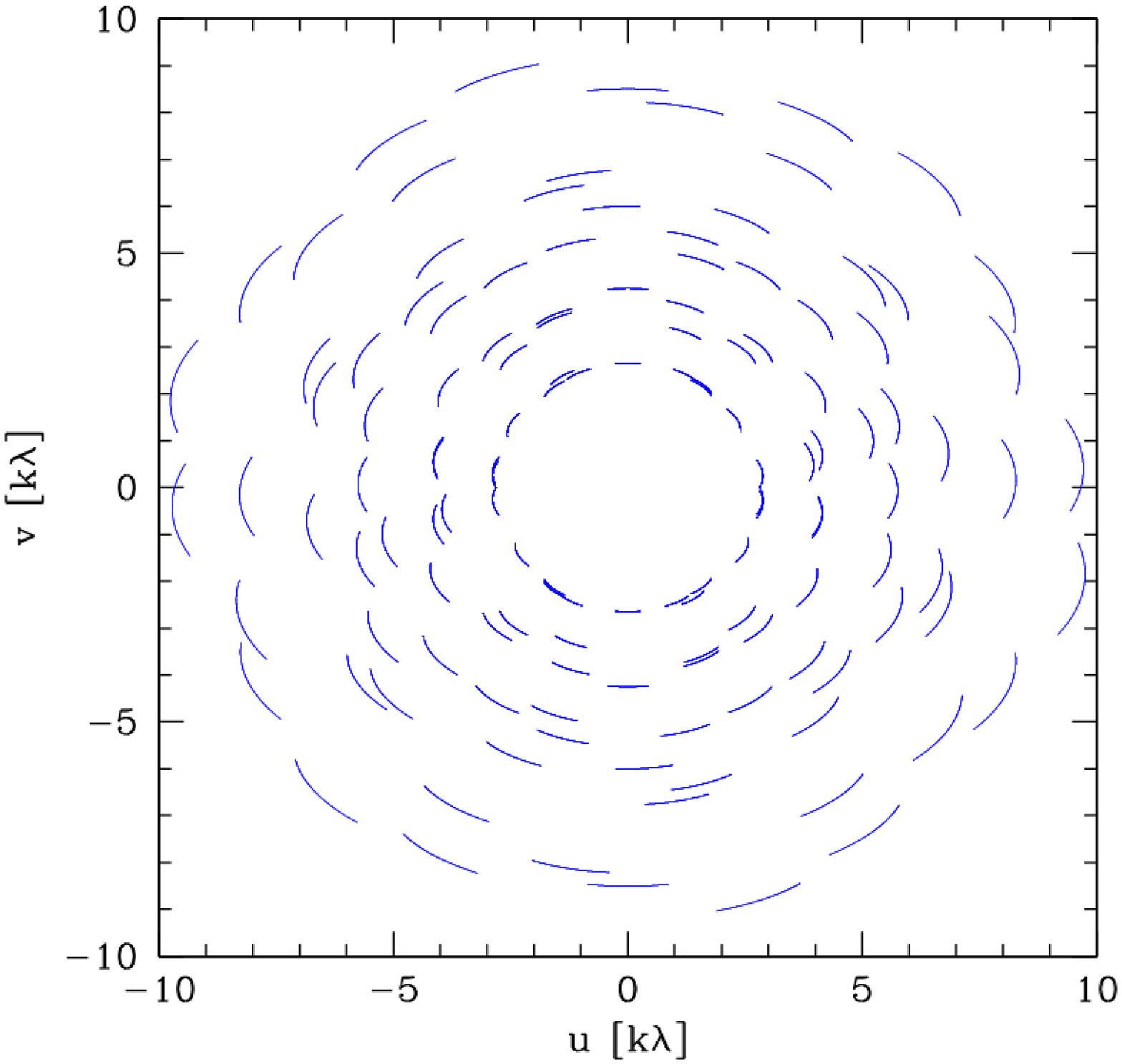}
   \FigureFile(70mm,50mm){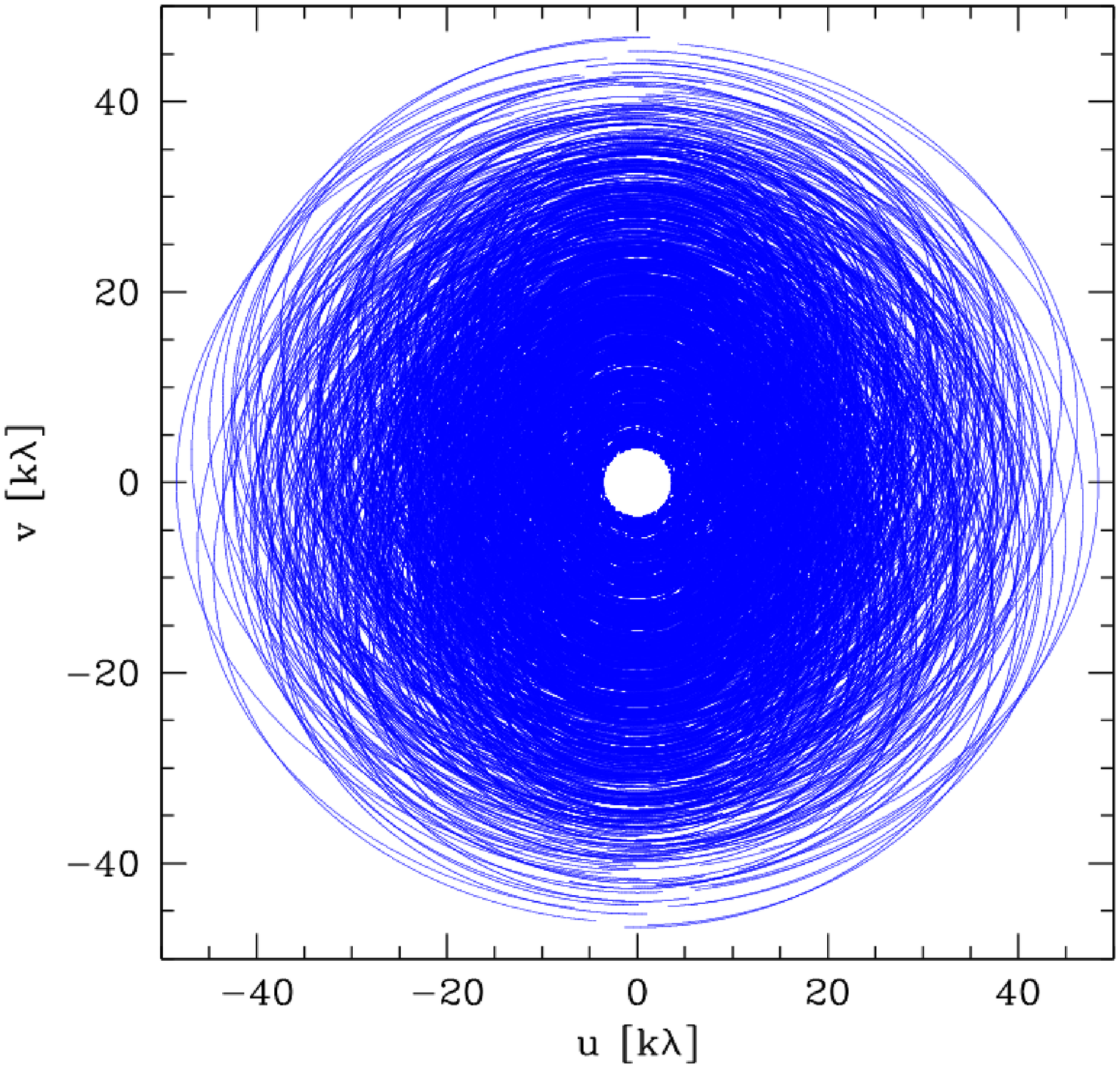}
   \FigureFile(70mm,50mm){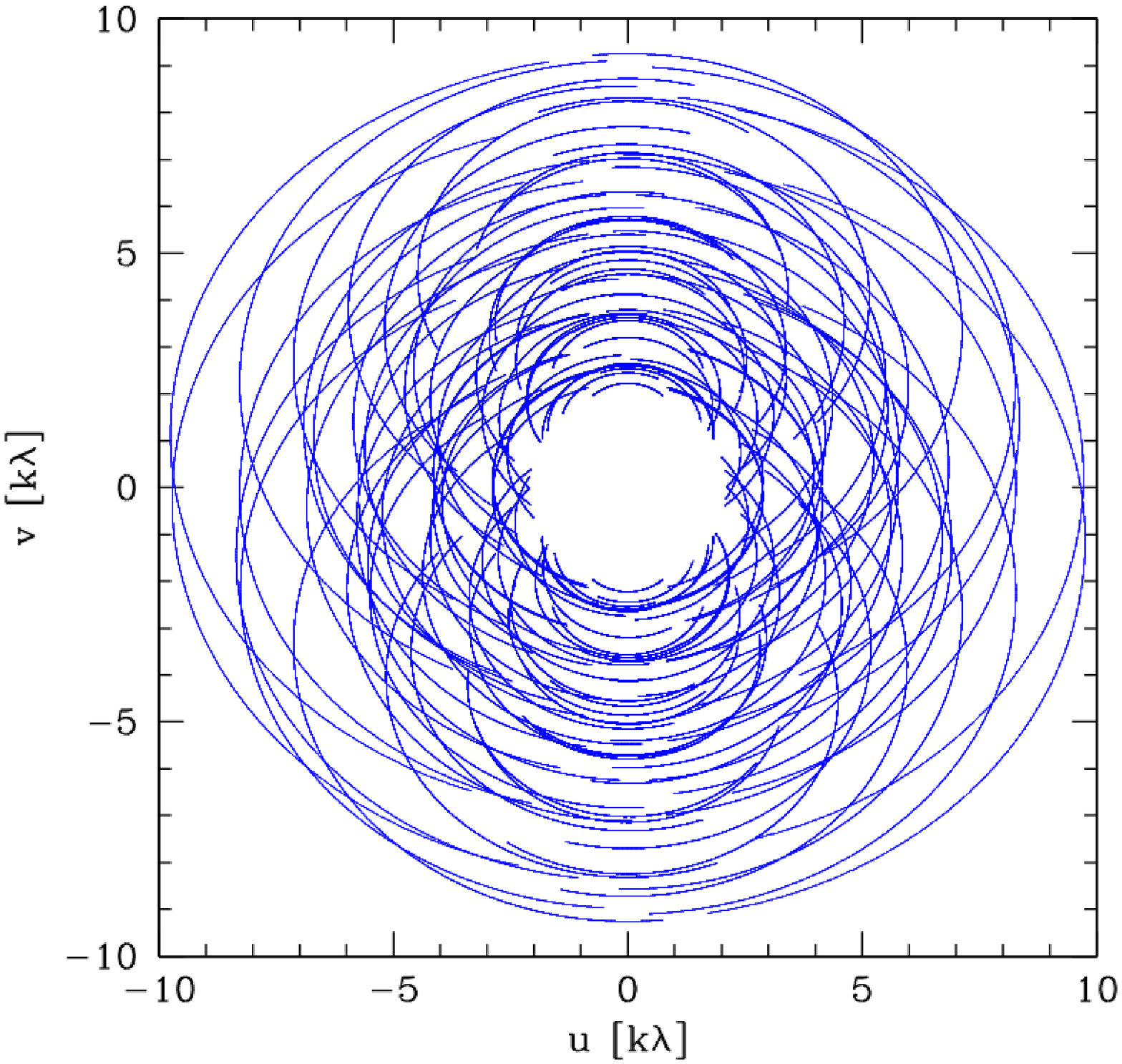}
   \FigureFile(70mm,50mm){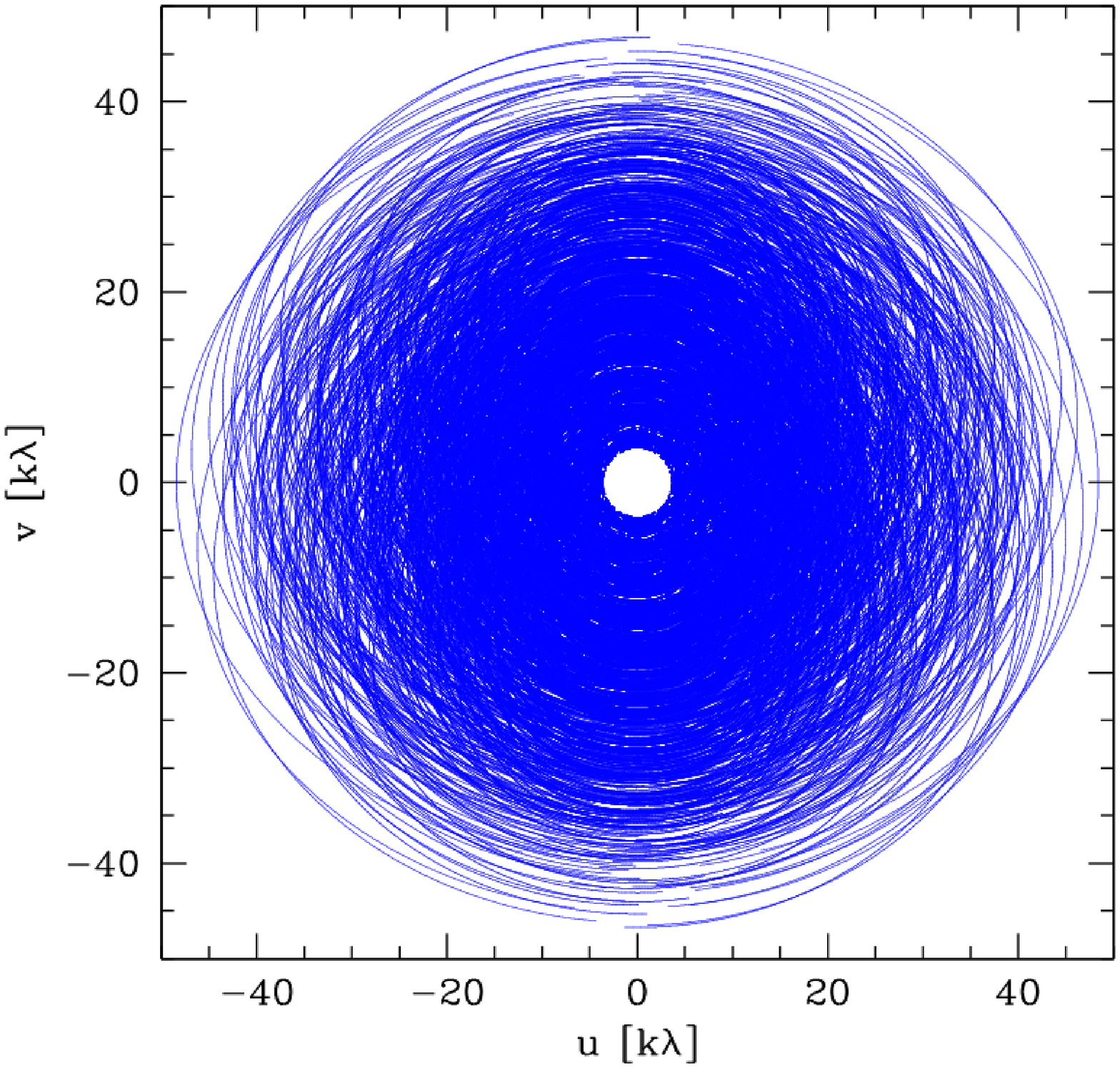}
   \FigureFile(70mm,50mm){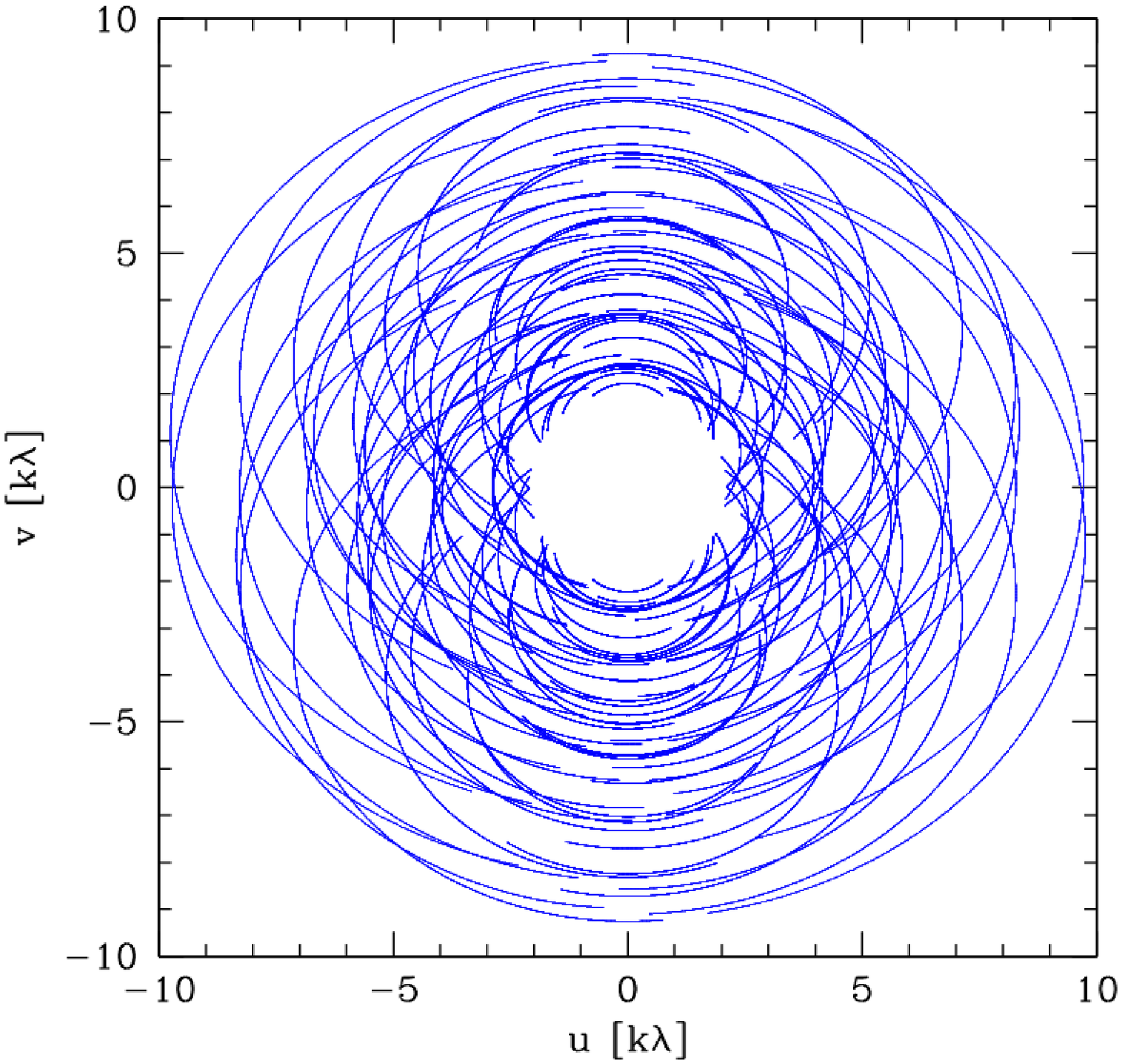}
  \end{center}
  \caption{The $u-v$ coverages of our mock observations with
 12m$\times$50 (left) and 7m$\times$12 (right) of Models A (top), B
 (middle), and C (bottom).    }  
\label{fig-uv}
\end{figure*}

In addition, single-dish images are created in real space assuming a
Gaussian beam with $69''$ FWHM and taking account of the thermal
noise. An additional factor of $\sqrt{2}$ is multiplied to the thermal
noise level so as to take account of beam switching. Since the spatial
resolution of the single-dish data is considerably worse than that of
12m$\times$50 and 7m$\times$12, the single-dish mapping is performed
over a larger sky area to reduce the edge effect; an extra $35''$ (0.5
FWHM) guard band is observed on each side of the image to cover
$190''\times 190''$ (Models A and C) or $310''\times 310''$ (Model
B). As the primary beams of 12m$\times$50 and 7m$\times$12 also extend
into the guard band, we will be able to detect point sources in this
band to some extent as will be shown in \S \ref{sec-bullet}.  The pixel
size of the single-dish image is taken to be $10''$ (Table
\ref{tab-models}).  While this somewhat oversamples the data, we adopt
this pixel size in the present analysis for the following reasons; i) as
the expected random noise of the single-dish image reduces linearly with
the pixel size, the above choice serves as a conservative limit, ii) the
thermal noise of the single-dish data in $\mu$Jy/arcsec$^2$ is already
smaller than that of 12m$\times$50 with the above pixel size (Table
\ref{tab-models}), and iii) we find that adopting larger pixel sizes
often results in spurious distortions in the reconstructed images after
deconvolved jointly with the interferometric data. \citet{Yen11} also
show that one may well reproduce the extended emission by adopting pixel
sizes that are sufficiently smaller than the primary beam of an
interferometer.  Effective integration time at each point of the
single-dish image, including the time spent for beam switching, is given
by
\begin{equation}
t_{\rm eff,SD} = 
\frac{t_{\rm int, SD}}{N_{\rm point,SD}},
\end{equation}
where $t_{\rm int, SD}$ is the total integration time and $N_{\rm
point,SD}$ is the number of pointings of SD$\times$4 listed in Table
\ref{tab-models}.

The mock data so obtained are reduced without a priori knowledge of the
input model or point sources as described in the following subsections.
The point source removal and image deconvolution are both performed over
the extended regions with the guard band mentioned above.  Impacts of
other sources of noise than considered here will be discussed in \S
\ref{sec-error}.

\subsection{Point source removal}
\label{sec-source}

Whenever point sources are present, we make use of long baseline
data to remove them before reconstructing the extended emission.
First, a dirty map and a dirty beam are created only for
12m$\times$50 by an inverse Fourier transformation of mock visibilities
using the MIRIAD task {\it invert}.

Second, the position and the apparent flux of a source whose peak
brightness lies above a certain threshold value are identified with the
CLEAN algorithm by \citet{Steer84} using the MIRIAD task {\it
mossdi}. If there are multiple sources, the flux of each source may be
mutually influenced via the side-lobe effects. A single point source may
also be detected as a superposition of separate sources. In either case,
the apparent flux of the $i$-th source can be expressed as
\begin{equation}
F^{\rm app}_{i} = \sum_{j=1}^{N} S_{ij} F^{\rm int}_{j},
\end{equation}\label{PS_int}
where $N$ is the total number of detections, $F^{\rm int}_{j}$ is the
intrinsic flux of the $j$-th source, and the matrix $S_{ij}$ describes
the contribution of $F^{\rm int}_{j}$ to $F^{\rm app}_{i}$ for a given
beam shape and source positions.  Given knowledge of the $u-v$ coverage
of the observations, one can estimate the components of $S_{ij}$ and
solve the linear algebraic equations for $F^{\rm int}_{j}$. In the
present analysis, all the detections that lie within the synthesized
beam FWHM from each other are regarded as components of a single source,
whose flux and position are given by the sum and the flux-weighted mean,
respectively, of fluxes and positions of all the components.

Finally, contribution of identified sources is removed from the
visibility data of both 12m$\times$50 and 7m$\times$12 as well as from
the SD$\times$4 image.

\subsection{Image deconvolution}

Having removed the identified point sources, we create dirty images and
dirty beams for either each of or the combinations of 12m$\times$50 and
7m$\times$12.  The Briggs robust weighting scheme \citep{Briggs95} is
adopted with the robustness parameter of 0.5.  The dirty images are then
deconvolved (corrected for the effects of the dirty beams and
re-convolved with a Gaussian beam) with the Maximum Entropy Method
(MEM), which is more suitable for the extended signals than CLEAN (e.g.,
\cite{Narayan86, Sault90}).  Non-linear deconvolution is applied using
the MIRIAD task {\it mosmem} to the dirty images either separately or
jointly with the single-dish image.  Since the SZE brightness is
negative at $\nu <220$ GHz while MEM is applicable only to the positive
signals, we change the sign of the entire maps when applying MEM. The
pixel size is taken to be the same as the input image ($1''$).

Since a proper estimate of the underlying noise level of an image with
an extended source is not straightforward, we adopt and compare the
following two definitions using the pixels within $120''\times 120''$
(Models A and C) or $240''\times 240''$ (Model B); 1) $\sigma_{\rm th}$
is the rms value of a deconvolved image with the same observing
parameters but using the {\it null} input model including only the
thermal noise, and 2) $\sigma_{\rm diff}$ is the rms value of the
difference map between the deconvolved output and the smoothed input;
the input described in \S \ref{sec-input} is convolved by a Gaussian
with the same FWHM as the synthesized beam of the output. The values of
$\sigma_{\rm th}$ are given in Table \ref{tab-models}. Thermal noise is
nearly constant over the map except near the edge.

Finally, we quantify the quality of reconstruction by the image
fidelity defined by (\cite{Pety01})
\begin{eqnarray}
f(\vec{\theta}) \equiv \frac{|I_{\rm in}^{\rm smooth}(\vec{\theta})|}
{\max\{|I_{\rm out}(\vec{\theta})-I_{\rm in}^{\rm smooth}(\vec{\theta})|,
~0.7 \sigma_{\rm diff}\}}
\end{eqnarray}
where $I_{\rm out}(\vec{\theta})$ is the output intensity of the deconvolved
image at a sky position $\vec{\theta}$ and $I_{\rm in}^{\rm smooth}
(\vec{\theta})$ is the input intensity smoothed with the same synthesized
Gaussian beam as the output. The above fidelity roughly corresponds
to an inverse of the fractional error of reconstruction. The second term
in the denominator is introduced to avoid an overshoot of the fidelity
due to a coincidental match between the input and the output.

\section{Results}
\label{sec-results}
\subsection{Model A: Gaussian}
\label{sec-gauss} 
\begin{figure*}[t]
  \begin{center}
   \FigureFile(75mm,50mm){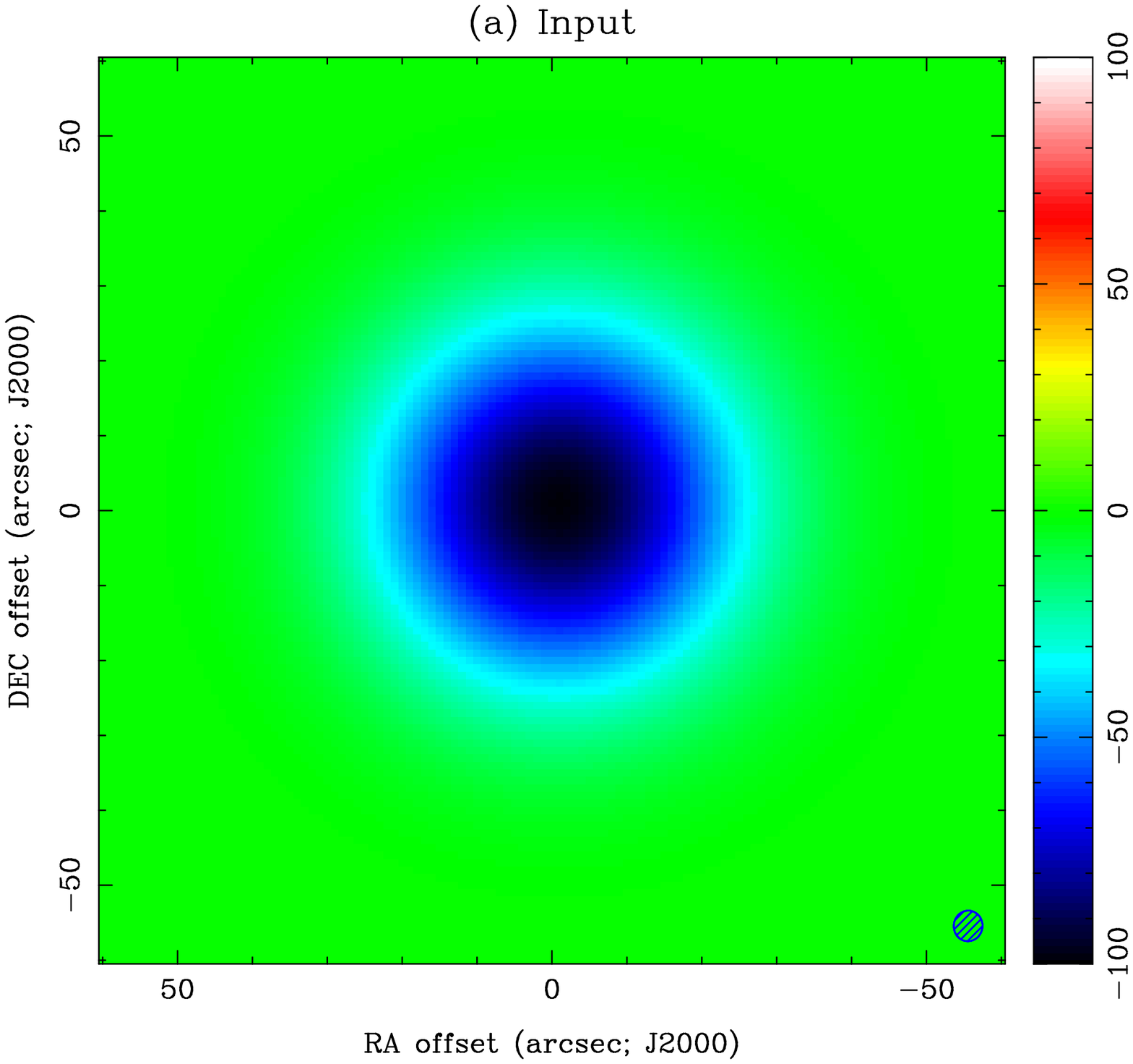}
   \FigureFile(75mm,50mm){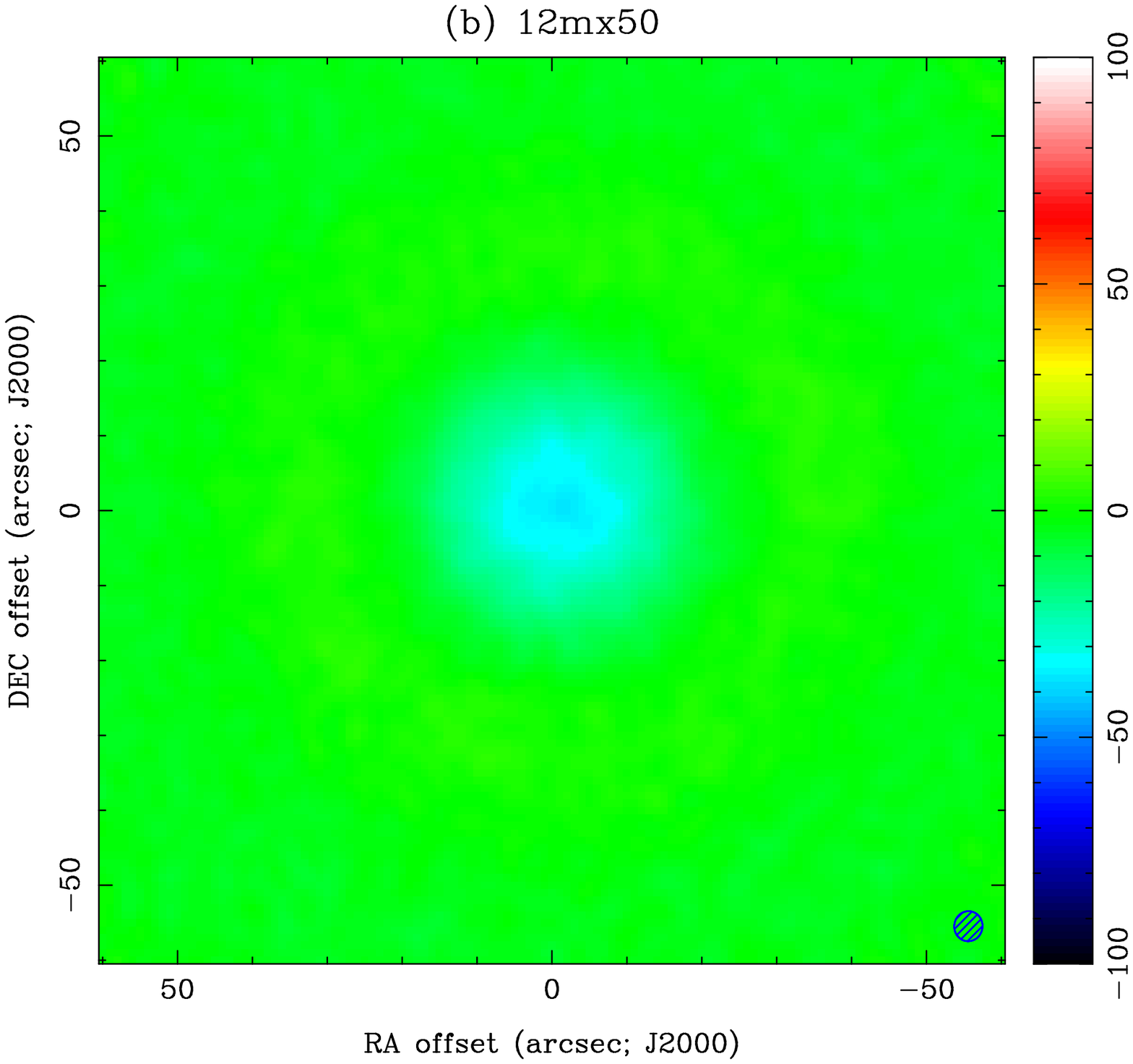}
   \FigureFile(75mm,50mm){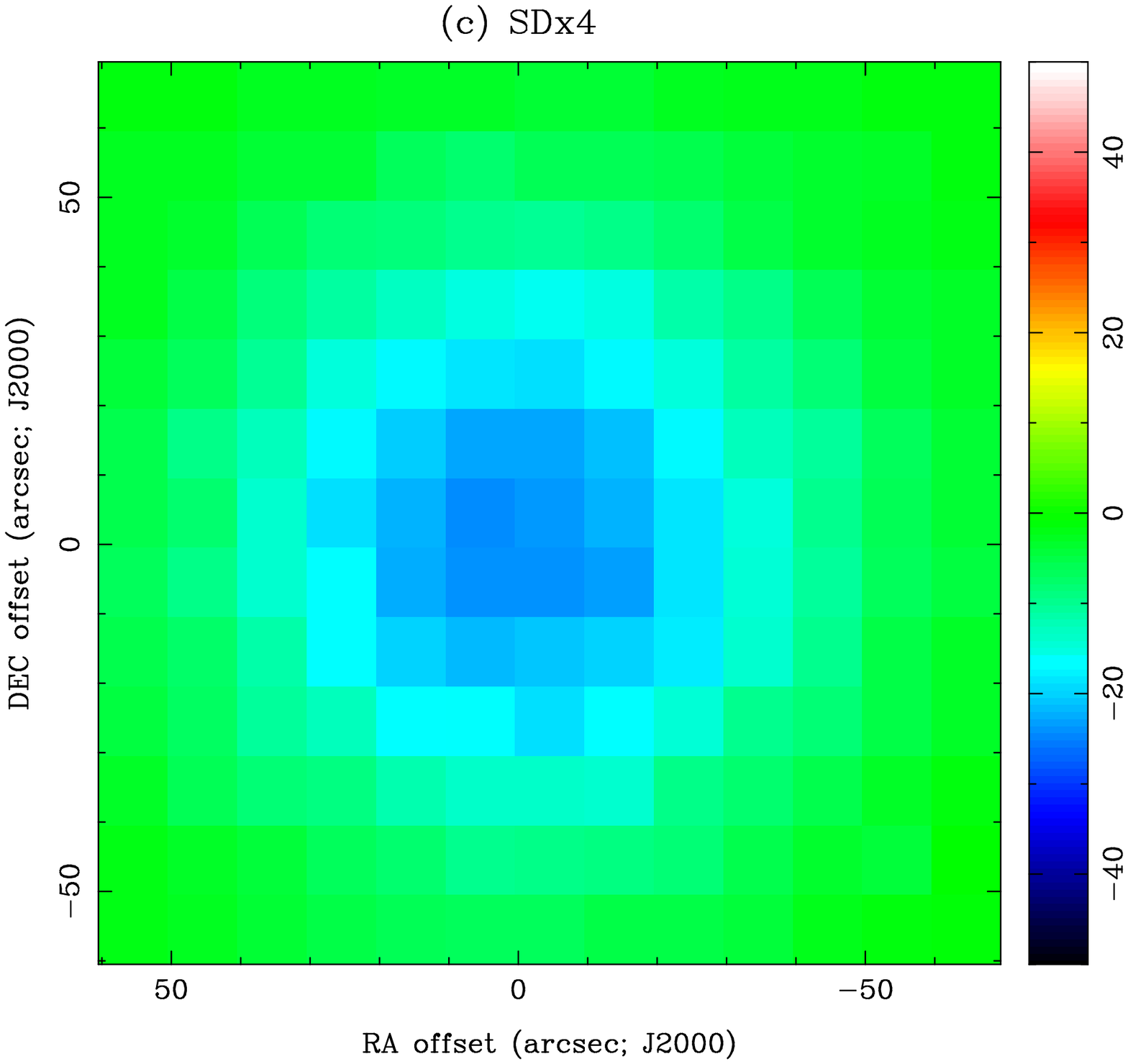}
   \FigureFile(75mm,50mm){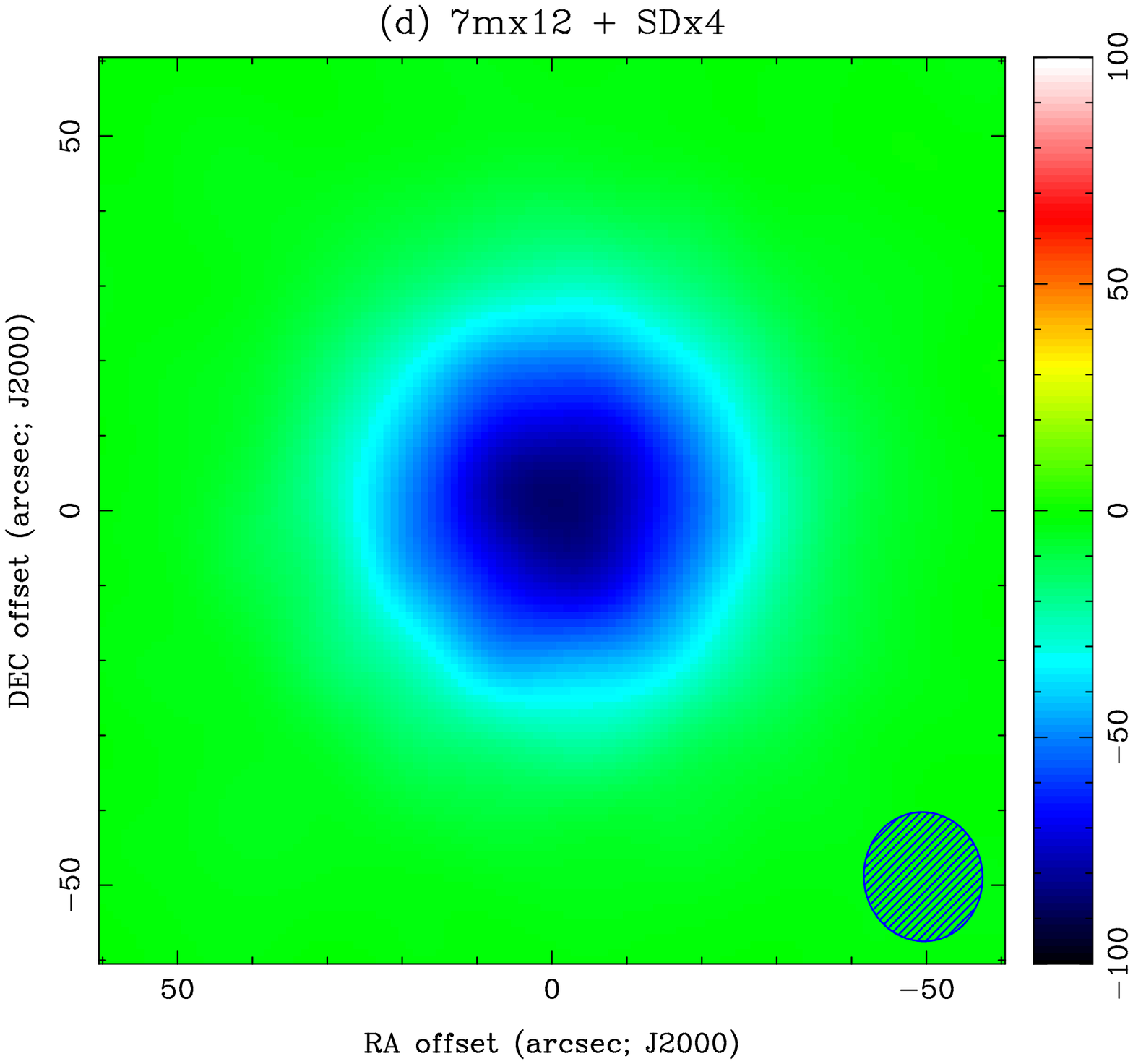}
   \FigureFile(75mm,50mm){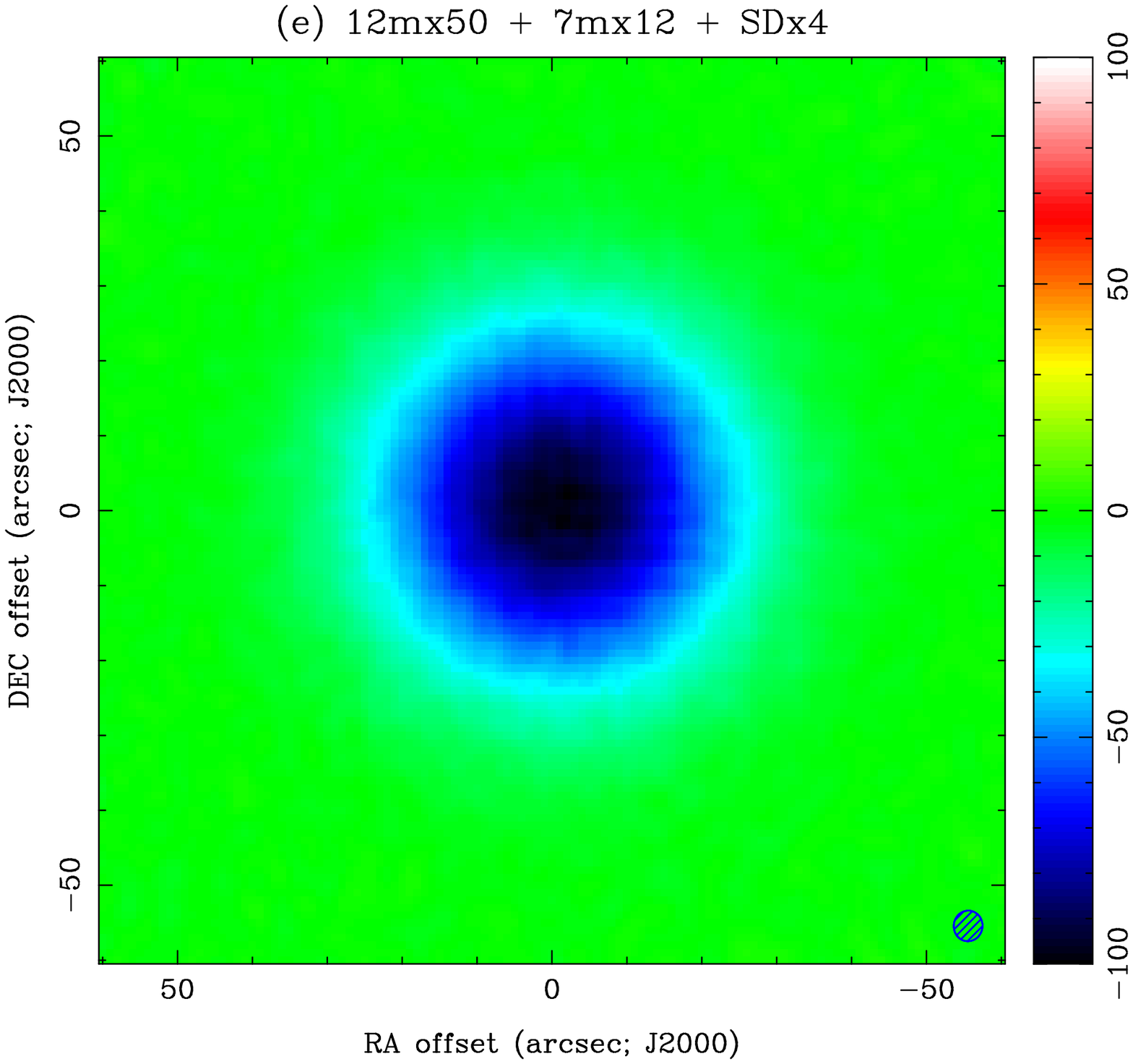}
    \FigureFile(75mm,50mm){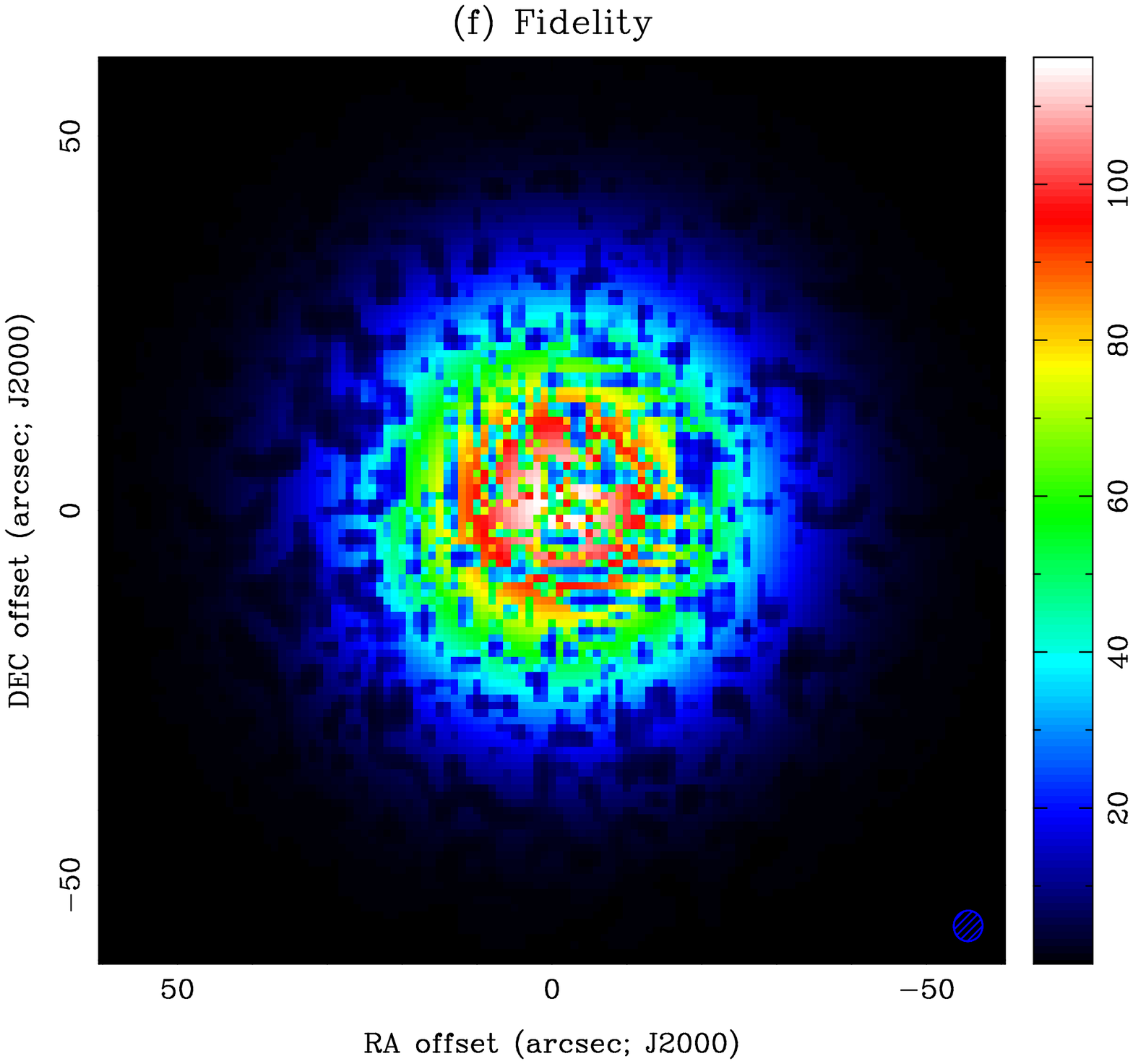}
  \end{center}
  \caption{Mock images of Model A (Gaussian) with $\theta_{\rm
model}=40''$.  Except for panel (f), the color scale is in
$\mu$Jy/arcsec$^2$.  A hatched ellipse at the bottom-right corner
indicates the synthesized beam. (a) Input model convolved with the same
synthesized beam as panel (e). (b) Deconvolved image for 12m$\times$50
with $(\sigma_{\rm th},\sigma_{\rm diff}) = (1.1, 17)$
[$\mu$Jy/arcsec$^2$] and $\theta_{\rm beam}=3.9''$.  (c) SD$\times$4
image with $\sigma_{\rm th} = 0.64$ $\mu$Jy/arcsec$^2$ and $\theta_{\rm
beam}=69''$.  (d) Deconvolved image for 7m$\times$12 + SD$\times$4 with
$(\sigma_{\rm th},\sigma_{\rm diff}) = (0.48, 0.79)$
[$\mu$Jy/arcsec$^2$] and $\theta_{\rm beam}=17''$.  (e) Deconvolved
image for 12m$\times$50 + 7m$\times$12 + SD$\times$4 with $(\sigma_{\rm
th},\sigma_{\rm diff}) = (1.0, 1.2)$ [$\mu$Jy/arcsec$^2$] and
$\theta_{\rm beam}=4.0''$.  (f) Fidelity of the image shown in panel
(e). }  \label{fig-gauss}
\end{figure*}

We plot in Figure \ref{fig-gauss} the results of a
representative case of a Gaussian model with $\theta_{\rm model}=40''$
FWHM after total integrations of 2 hr 12m$\times$50, 7m$\times$12, and
SD$\times$4, respectively. With the 12m$\times$50 data alone,
reconstruction is poor and extended ($\gg$ arcsec) feature is largely
missed even though the noise level of the 12m$\times$50 image (panel b)
is $\sim 1/100$ of the peak signal.  This is also indicated by the fact
that $\sigma_{\rm diff} \gg \sigma_{\rm th}$.  On the other hand,
significant improvement in the image fidelity is achieved once low
spatial frequency data from 7m$\times$12 and SD$\times$4 are included
(panels d, e, f). The brightness of the SD$\times$4 image appears to be
low simply because it is diluted by a large beam (panel c).

\begin{figure}[t]
  \begin{center}
    \FigureFile(80mm,50mm){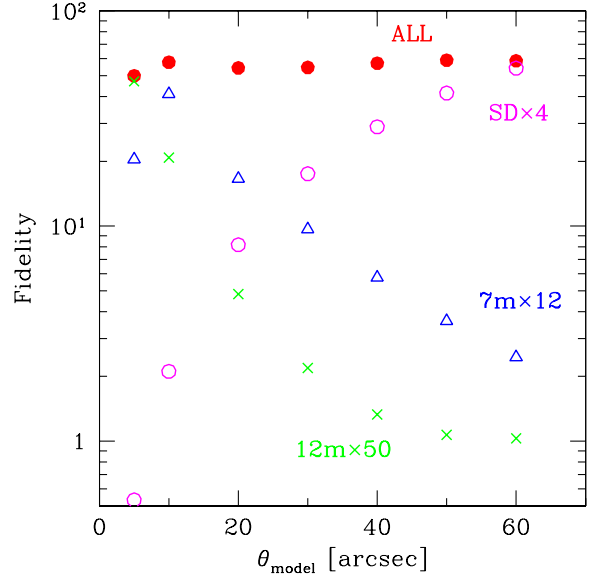}
  \end{center}
  \caption{Mean fidelity of the deconvolved images as a function of
  $\theta_{\rm model}$ for Model A. The mean is taken over the central
  $70''\times 70''$ and $\theta_{\rm model}\times \theta_{\rm model}$ for
  SD$\times$4 and the others, respectively. Symbols indicate the results
  for 12m$\times$50 (crosses), 7m$\times$12 (open triangles),
  SD$\times$4 (open circles), and 12m$\times$50 + 7m$\times$12 +
  SD$\times$4 (filled circles).}  \label{fig-gaussfidel}
\end{figure}

Figure \ref{fig-gaussfidel} illustrates how the mean fidelity around the
center varies with the spatial extent of the emission.  It is evident
that the interferometric data lose sensitivity for the emission much
more extended than the size of their synthesized beams, i.e.,
$\theta_{\rm beam}\sim 5''$ for 12m$\times$50 and $\theta_{\rm beam}\sim
20''$ for 7m$\times$12.  The single-dish data play an essential role in
improving the sensitivity at larger scales.

\begin{figure*}[t]
  \begin{center}
    \FigureFile(75mm,50mm){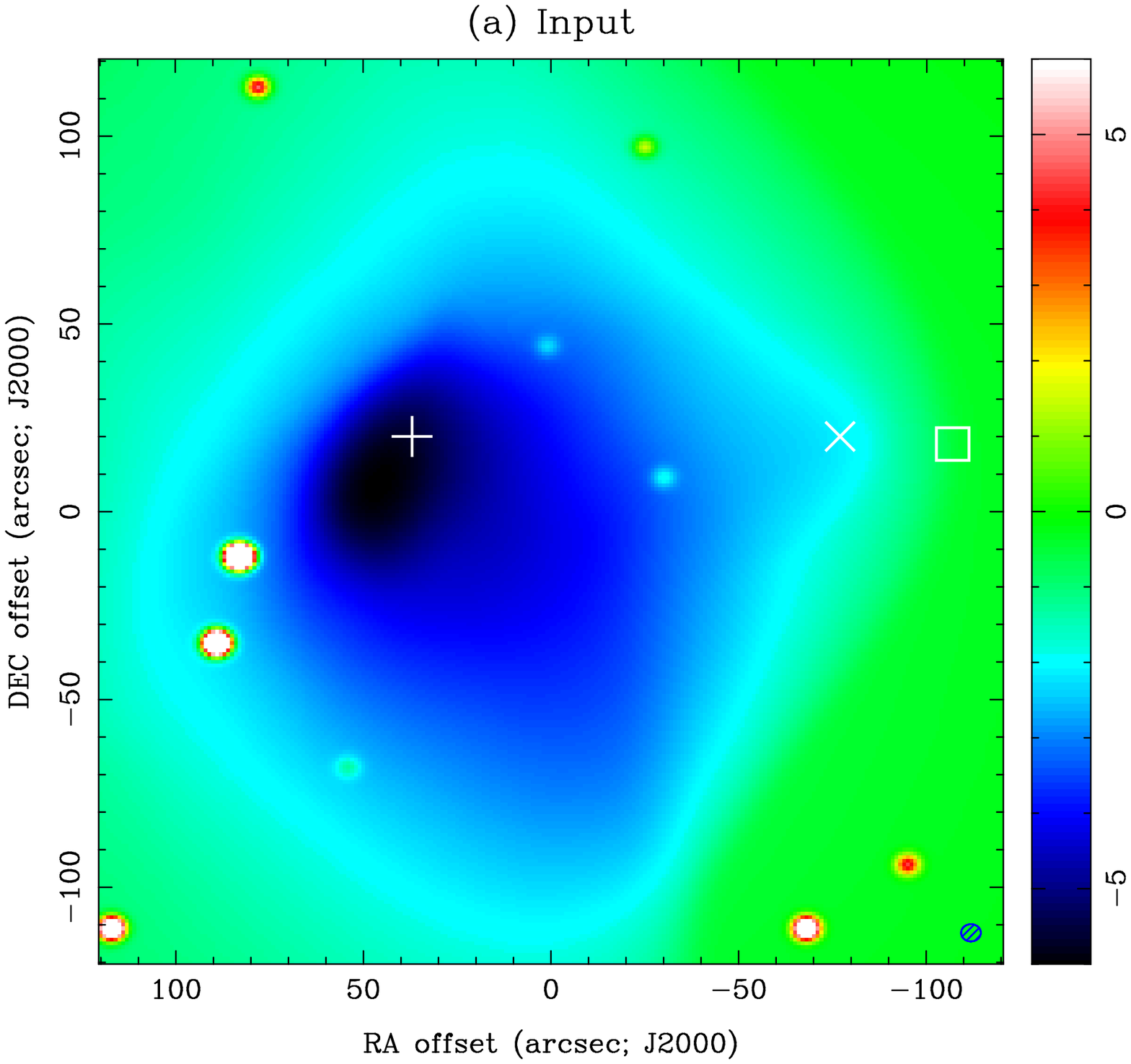}
    \FigureFile(75mm,50mm){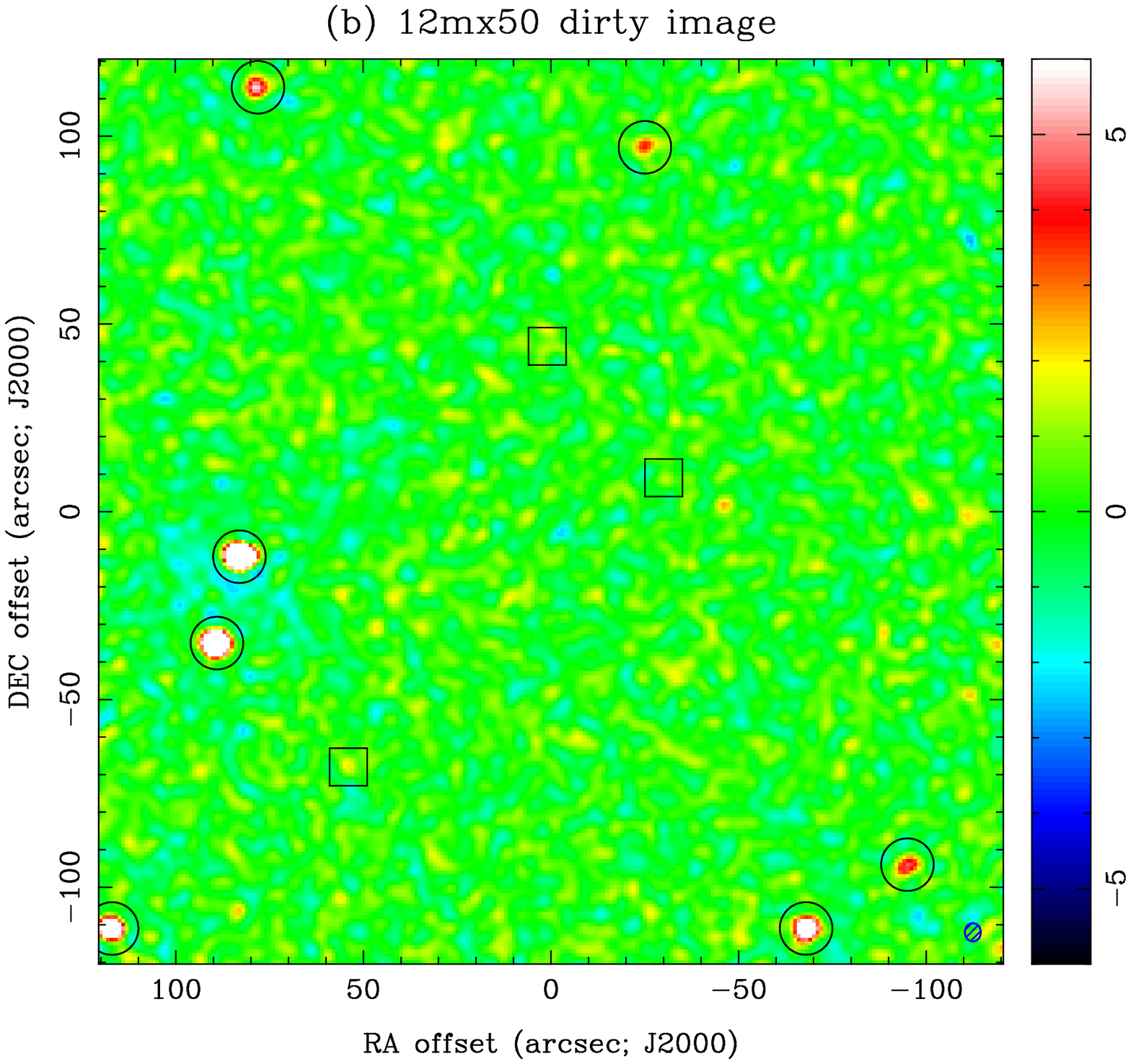}
    \FigureFile(75mm,50mm){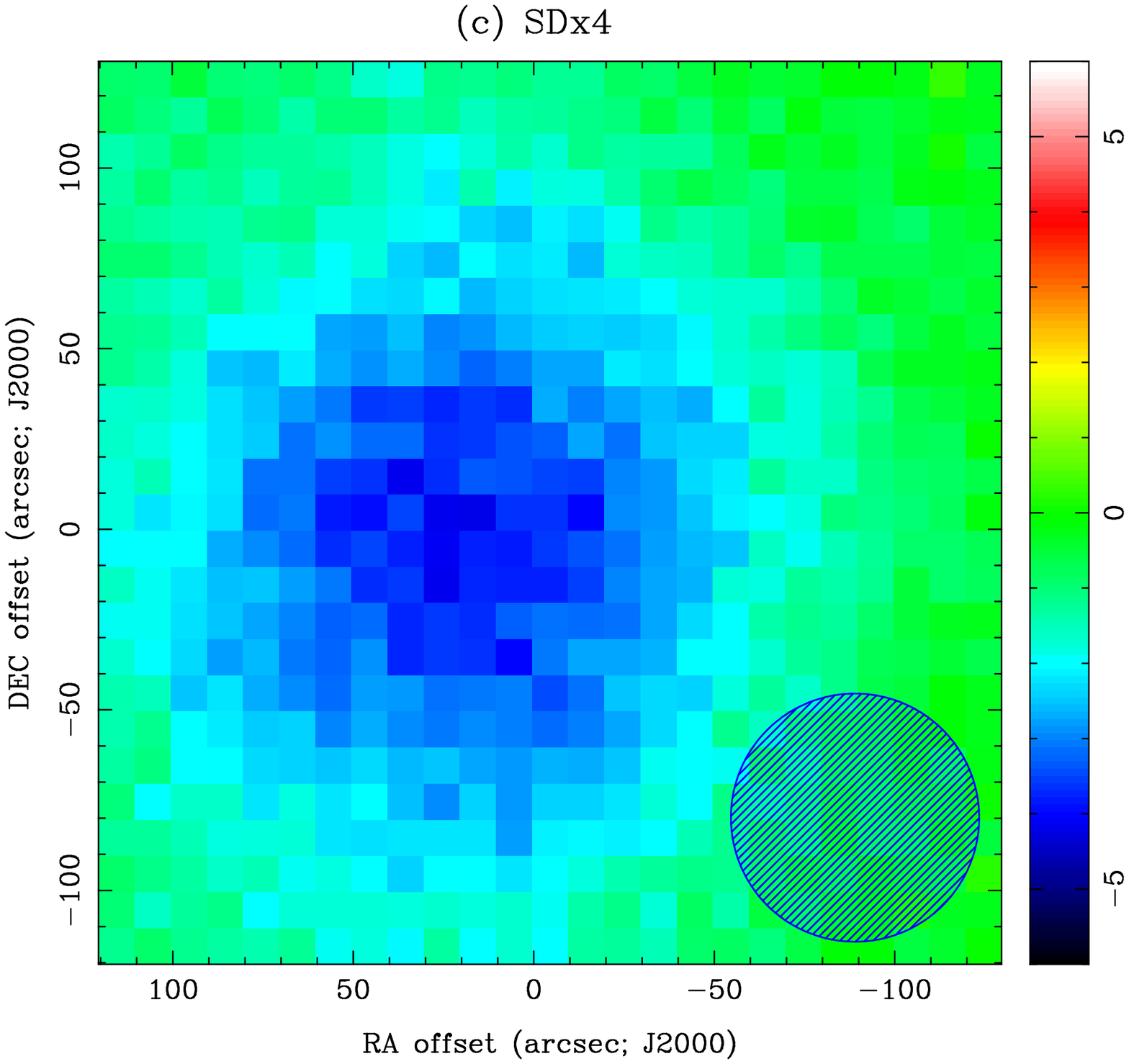}
    \FigureFile(75mm,50mm){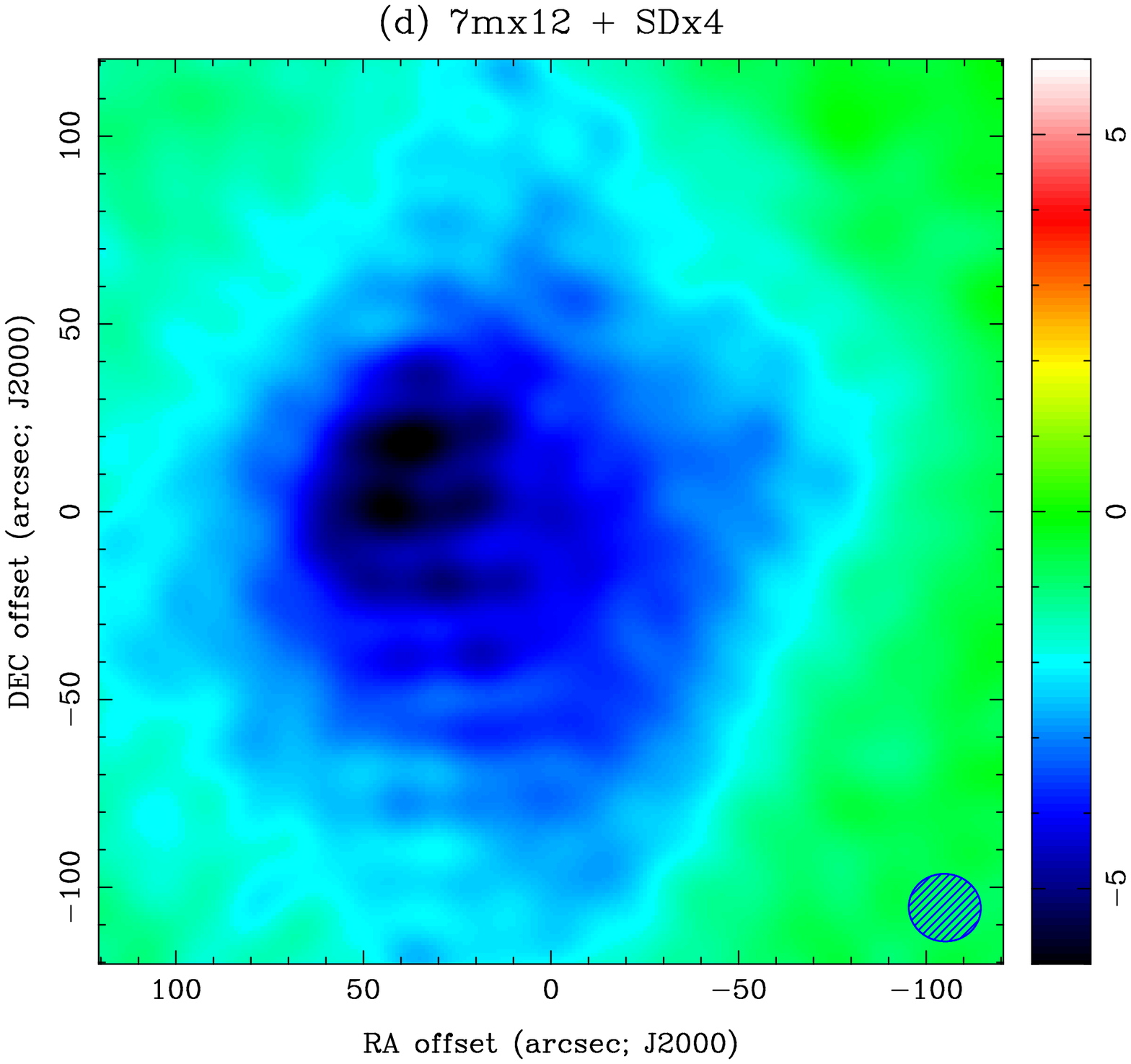}
    \FigureFile(75mm,50mm){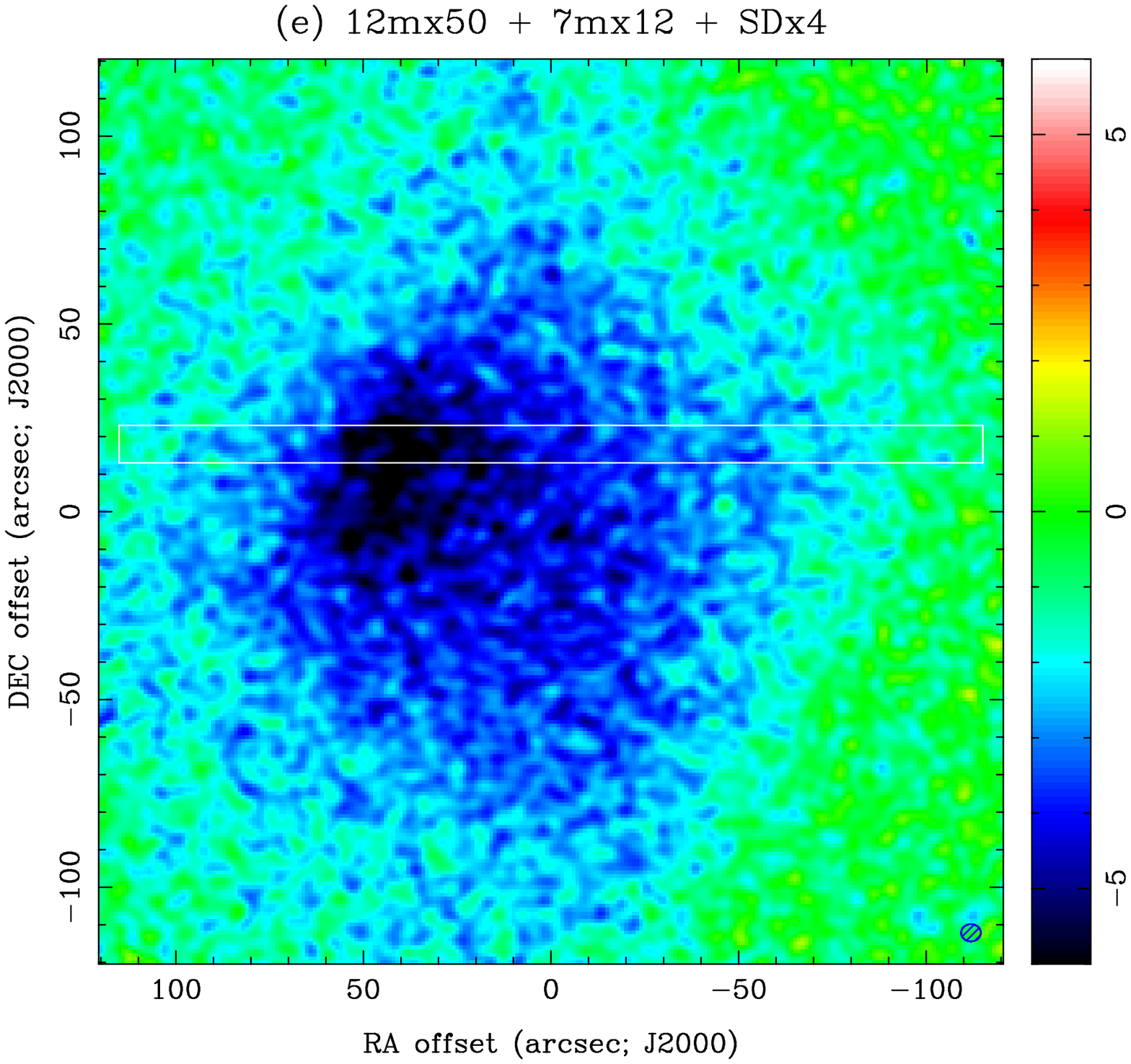}
    \FigureFile(75mm,50mm){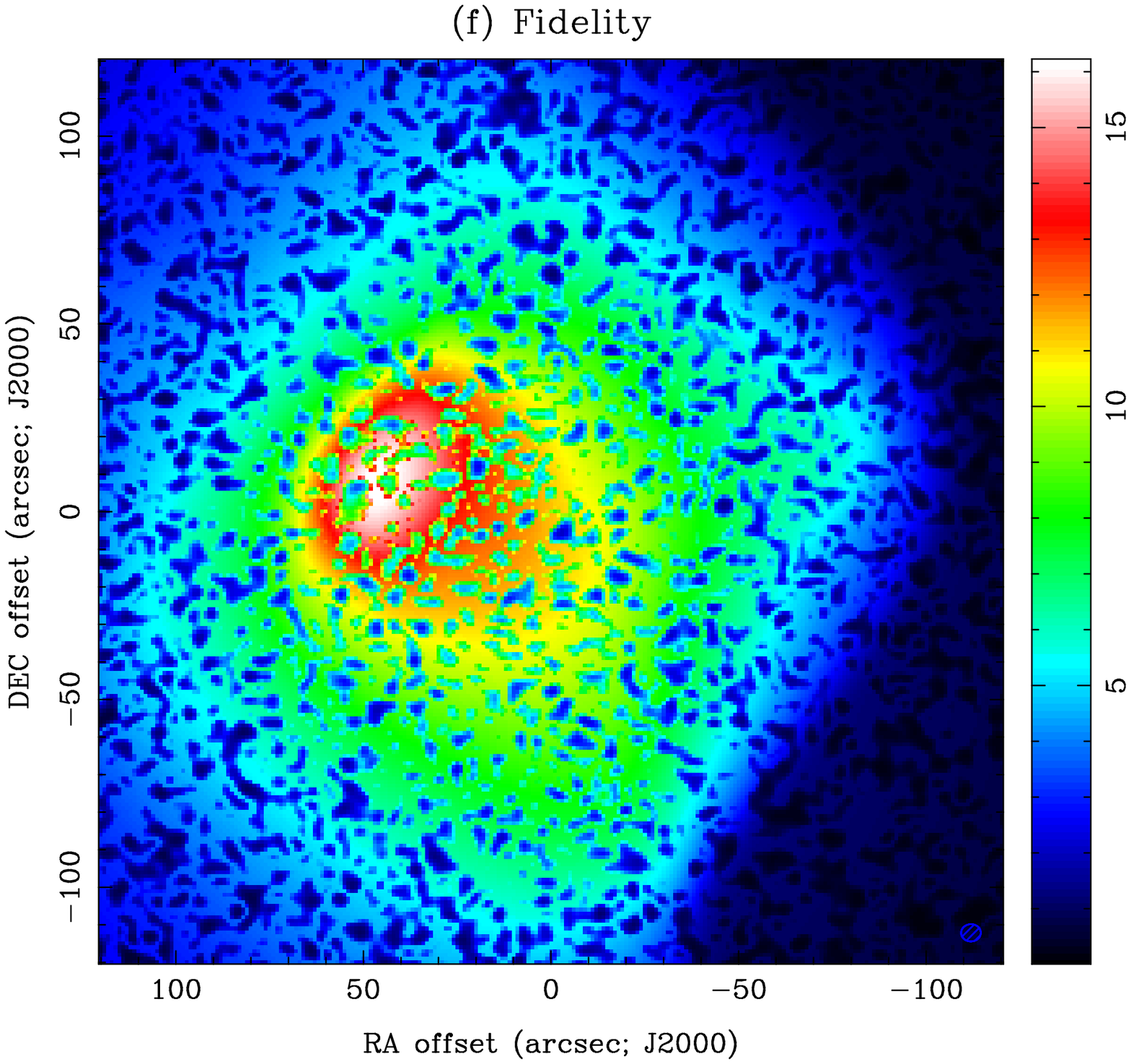}
  \end{center}
  \caption{Similar to Fig.~\ref{fig-gauss} but for Model B (bullet
cluster).  (a) Input model convolved with the same synthesized beam as
panel (e). Symbols indicate the positions of the X-ray peaks of the main
cluster (plus) and the sub-cluster (cross), and the shock front
(square). (b) Dirty image of 12m$\times$50 with $\sigma_{\rm th} = 0.62
~\mu$Jy/arcsec$^2$ and $\theta_{\rm beam}=4.5''$. Circles indicate the
point sources that are detected and removed in the subsequent panels,
whereas squares denote undetected ones.  (c) SD$\times$4 image with
$\sigma_{\rm th} = 0.23~\mu$Jy/arcsec$^2$ and $\theta_{\rm beam}=69''$.
(d) Deconvolved image for 7m$\times$12 + SD$\times$4 with $(\sigma_{\rm
th},\sigma_{\rm diff}) = (0.15 , 0.18)$ [$\mu$Jy/arcsec$^2$] and
$\theta_{\rm beam}=19''$.  (e) Deconvolved image for 12m$\times$50 +
7m$\times$12 + SD$\times$4 with $(\sigma_{\rm th},\sigma_{\rm diff}) =
(0.52, 0.53)$ [$\mu$Jy/arcsec$^2$] and $\theta_{\rm beam}=4.9''$. A box
indicates the region over which the profile in Fig.~\ref{fig-bulletsect}
is computed.  (f) Fidelity of the image shown in panel (e).}
\label{fig-bullet}
\end{figure*}

\begin{figure}[t]
  \begin{center}
    \FigureFile(80mm,50mm){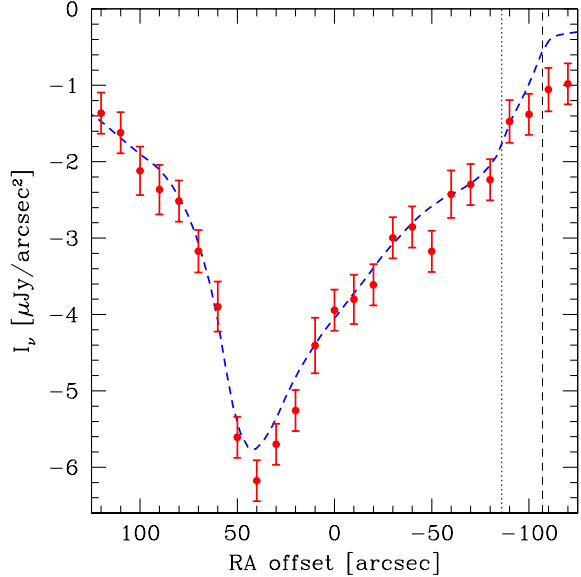}
  \end{center}
  \caption{Comparison of the input model (thick dashed line) and the
  mock data (error bars) for Model B over a strip with width $10''$
  shown in Figure \ref{fig-bullet}(e).  The bin size is $10''$.  The
  vertical lines indicate the positions of the shock front (thin dashed)
  and the contact discontinuity (thin dotted), respectively, on the
  collision plane of the two merging clusters.}  \label{fig-bulletsect}
\end{figure}

\subsection{Model B: Bullet cluster}
\label{sec-bullet}

Figure \ref{fig-bullet} shows the images of a simulated bullet cluster
after total integrations of 10 hr for 12m$\times$50 and 40 hr for
7m$\times$12 and SD$\times$4. We have fiducially fixed the ratio between
the integration times at its nominal value of 1:4 \citep{Pety01,
Morita05}. The impact of changing the ratio will be discussed in \S
\ref{sec-reltime}.

The morphology of the input SZE image (Fig.~\ref{fig-bullet}a) is quite
different from that in X-rays plotted in Figure 1 of
\citet{Akahori12}. In particular, the so-called ``bullet'', the X-ray
peak of the sub-cluster, is not apparent in the SZE image as it lies
near the contact discontinuity across which pressure is nearly
constant. On the other hand, the shock front, i.e., the pressure gap,
ahead of the bullet is more prominent (see also Figs
\ref{fig-bulletsect} and \ref{fig-shocksect}) and extends over several
hundred kpc. The SZE peak lies near the X-ray peak of the main cluster
where electron density and temperature are both high.

There are 12 radio point sources over $310'' \times 310''$ among which
10 lie within the $240'' \times 240''$ region mapped by 12m$\times$50
(Fig.~\ref{fig-bullet}a). All the sources brighter than $\sim$0.1 mJy
are detected above a conservative threshold of 7 $\sigma_{\rm th}$ in
the 12m$\times$50 data as indicated in Table \ref{tab-sources} and
Figure \ref{fig-bullet}(b). After correcting for the side-lobe effects
as described in \S \ref{sec-source}, reconstructed fluxes agree with the
input values within $20\%$ for the resolved point sources and the source
positions are identified more accurately.  While the aperture bias of
the source fluxes has been enhanced by pointing errors, it is still
comparable to the thermal noise and we do not attempt to correct it
further in the present paper.  Two bright sources in the guard band are
also detected, although the noise level there is larger due to the
primary beam attenuation beyond the map edge. The remaining 3 fainter
sources are undetected but their contamination to the SZE is negligible.

Once the detected sources are removed, the deconvolved image by
12m$\times$50 + 7m$\times$12 + SD$\times$4 reproduces the input model
with the maximum fidelity reaching $\sim$ 16 (panels e and f in
Fig.~\ref{fig-bullet}). The fact that $\sigma_{\rm th} \simeq
\sigma_{\rm diff}$ holds also assures that the reconstruction of the
extended signal is as good as expected.

Figure \ref{fig-bulletsect} further illustrates one-dimensional SZE
intensity profile across the shock front. The error bars indicate the
statistical error of the mean in each bin estimated by
\begin{equation}
\sigma_{\rm bin} = \frac{1}{\sqrt{N_{\rm beam}}}
\max\left( \sigma_{\rm std}, ~\sigma_{\rm th} \right),
\label{eq-sigmabin} 
\end{equation}
where $\sigma_{\rm std}$ is the standard deviation of the pixel data in
the bin and $N_{\rm beam} = A_{\rm bin}/A_{\rm beam}$ is the number of
synthesized beams contained in the bin area $A_{\rm bin}$.  In equation
(\ref{eq-sigmabin}), we use $N_{\rm beam}$, instead of the number of
pixels, because the pixel values in a deconvolved image are correlated
over the beam area $A_{\rm beam}$. Note also that $\sigma_{\rm bin}$
does not include explicitly the thermal noise of the single-dish data
(see Table \ref{tab-models}) that is responsible for fluctuations on
spatial scales larger than $A_{\rm bin}$. It is evident from this figure
that the overall structure of this cluster is well reconstructed,
although the shock front is only marginally resolved.  We therefore
explore the feasibility of a targeted observation toward the shock
below.

\subsection{Model C: Shock front}
\label{sec-shock}

\begin{figure*}[t]
  \begin{center}
    \FigureFile(75mm,50mm){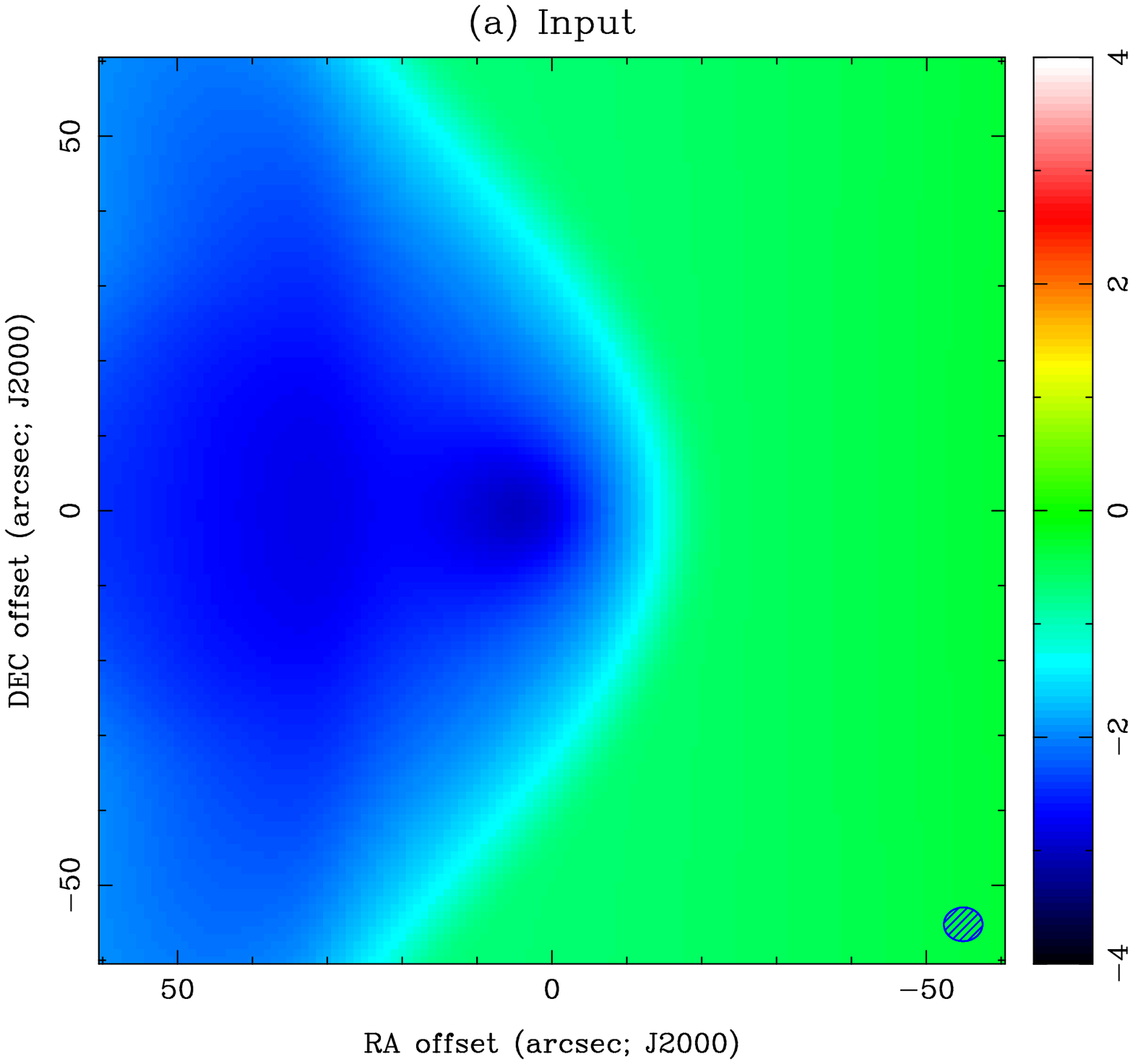}
    \FigureFile(75mm,50mm){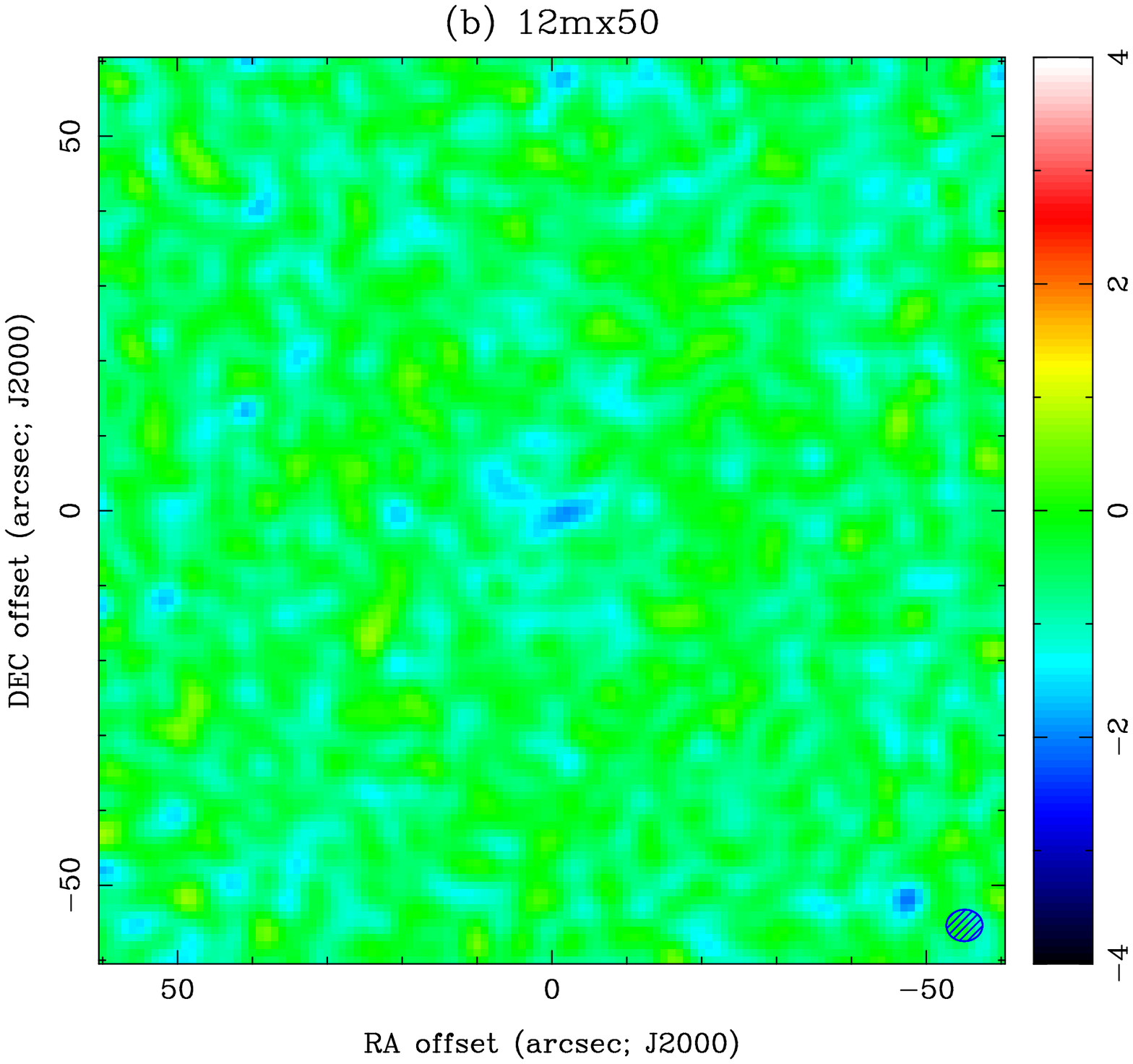}
    \FigureFile(75mm,50mm){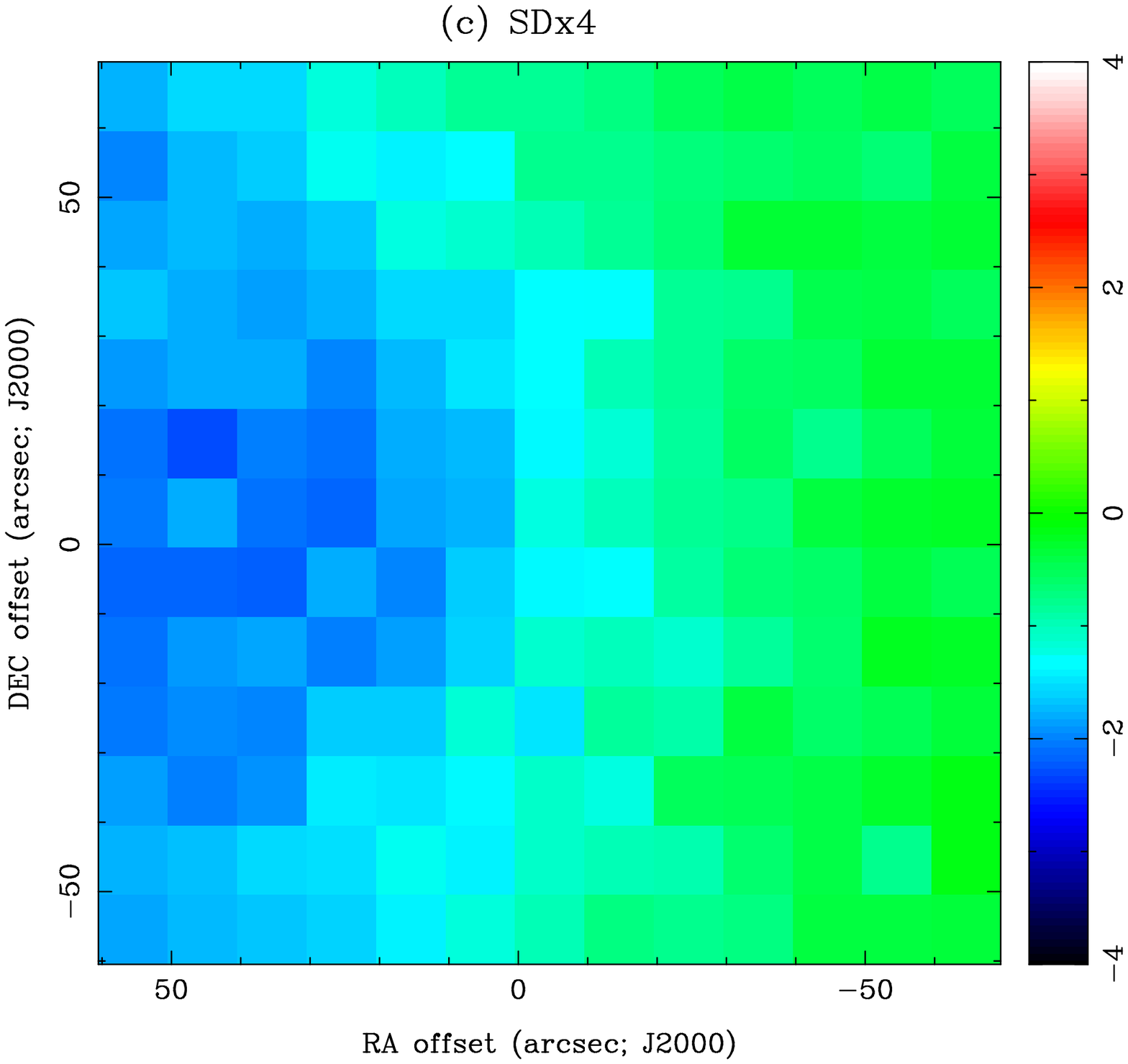}
    \FigureFile(75mm,50mm){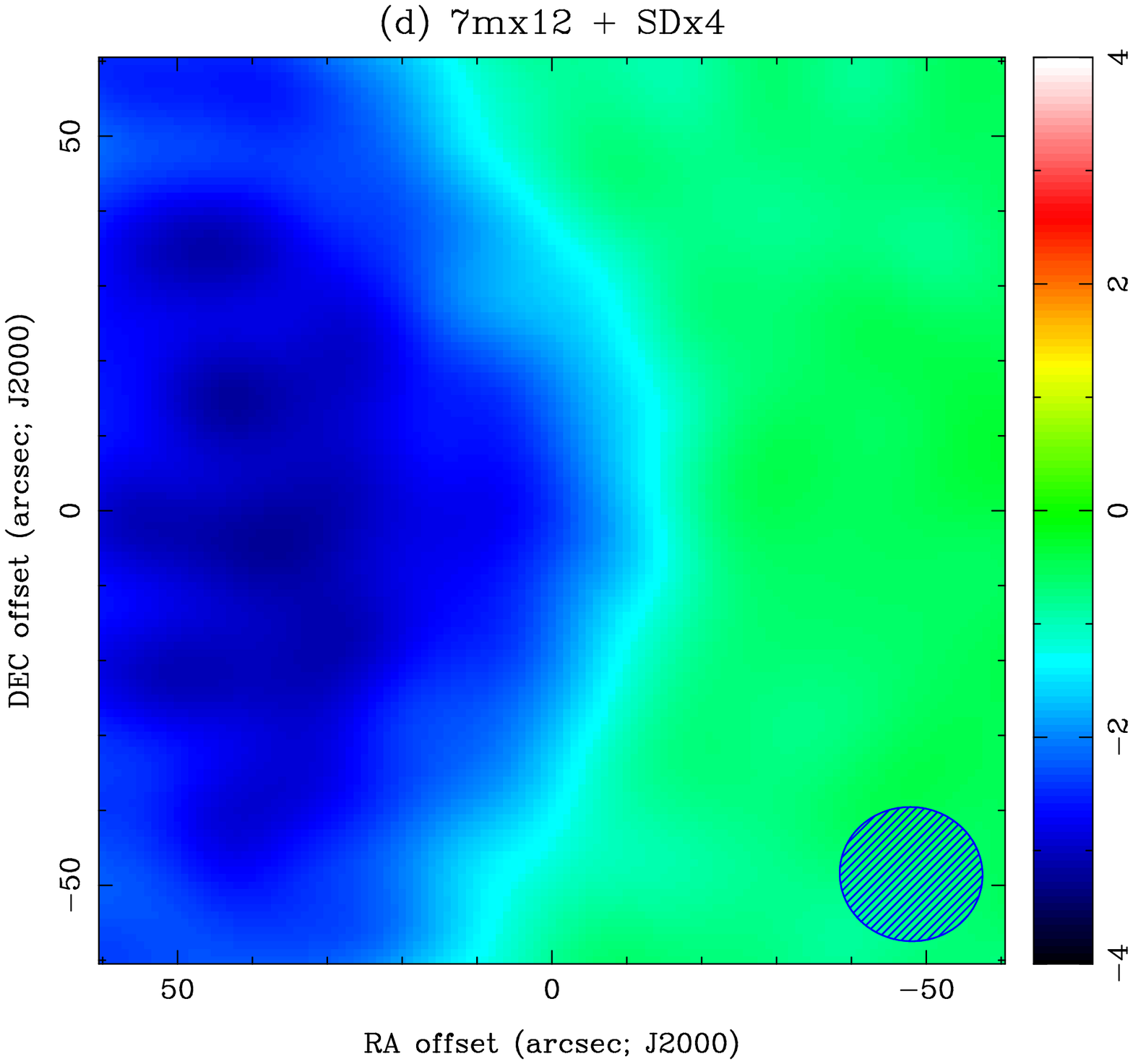}
    \FigureFile(75mm,50mm){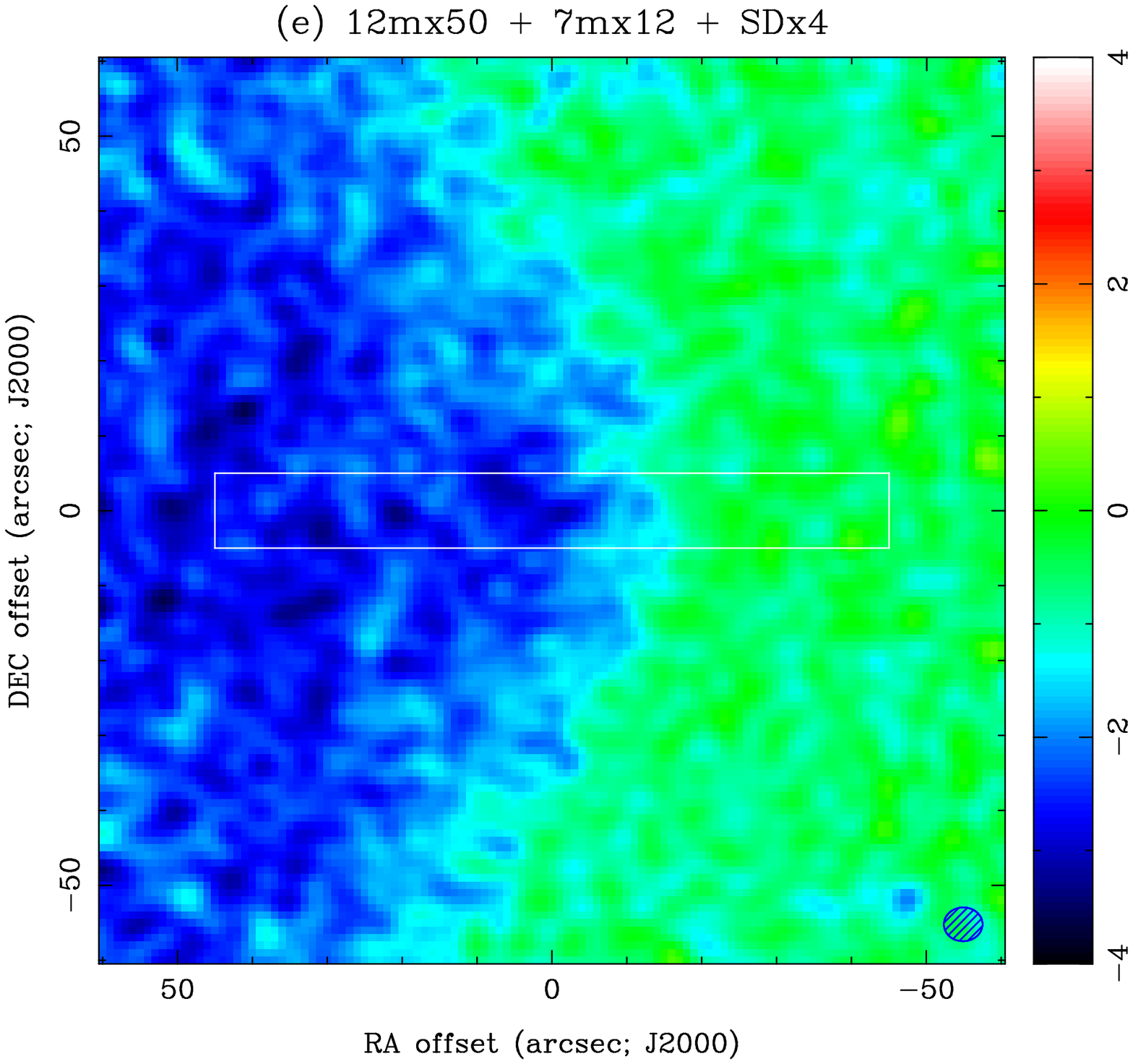}
    \FigureFile(75mm,50mm){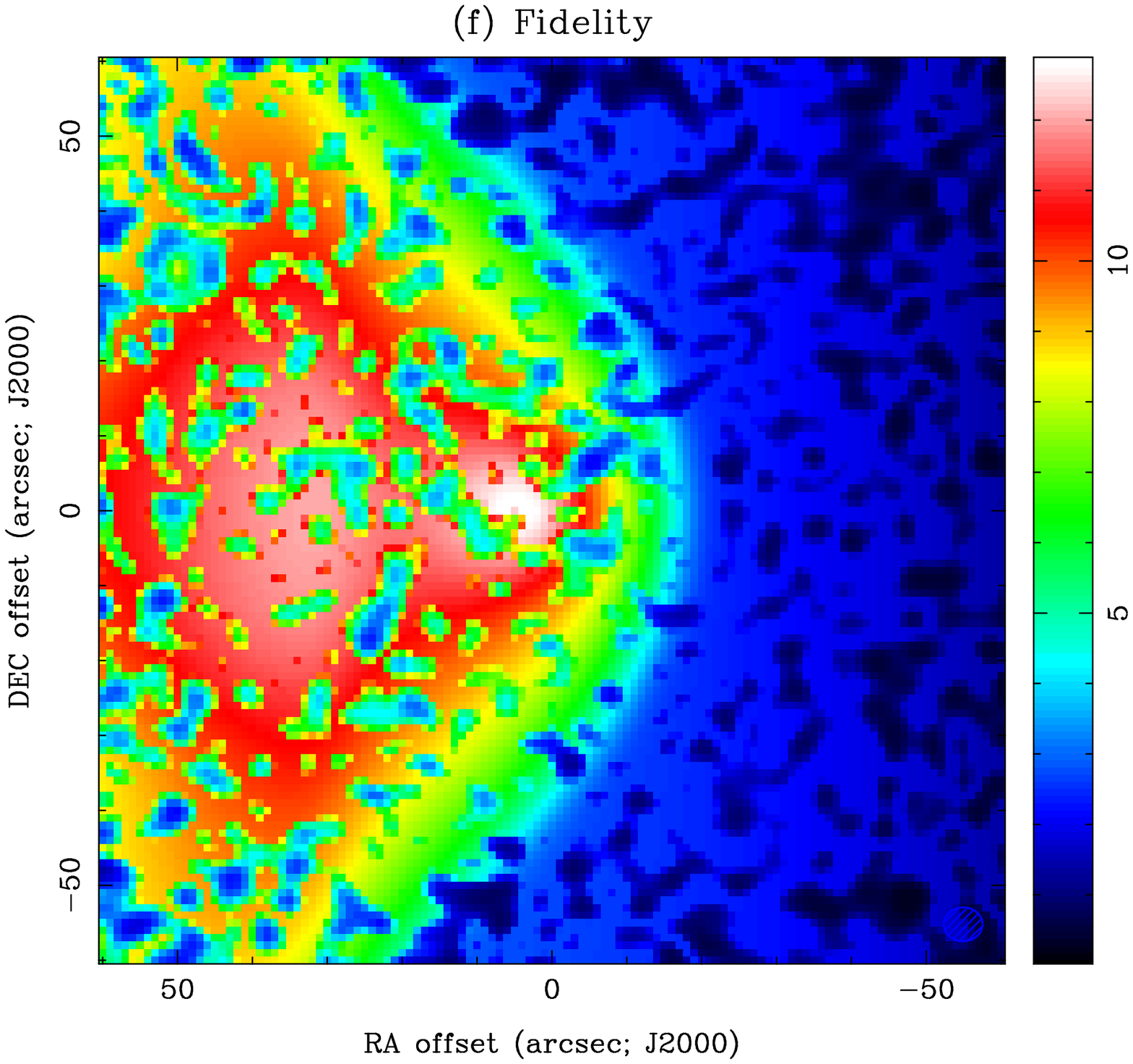}
  \end{center}
  \caption{Similar to Fig.~\ref{fig-gauss} but for Model C (shock
front). (a) Input model convolved with the same synthesized beam as
panel (e).  (b) Deconvolved image for 12m$\times$50 with $(\sigma_{\rm
th},\sigma_{\rm diff}), = (0.35, 0.95)$ [$\mu$Jy/arcsec$^2$] and
$\theta_{\rm beam}=4.5''$.  (c) SD$\times$4 image with $\sigma_{\rm th}
= 0.14$ $\mu$Jy/arcsec$^2$ and $\theta_{\rm beam}=69''$.  (d)
Deconvolved image for 7m$\times$12 + SD$\times$4 with $(\sigma_{\rm
th},\sigma_{\rm diff}), = (0.099, 0.18)$ [$\mu$Jy/arcsec$^2$] and
$\theta_{\rm beam}=18''$.  (e) Deconvolved image for 12m$\times$50 +
7m$\times$12 + SD$\times$4 with $(\sigma_{\rm th},\sigma_{\rm diff}) =
(0.30, 0.33)$ [$\mu$Jy/arcsec$^2$] and $\theta_{\rm beam}=4.8''$.  A box
indicates the region over which the profile in Fig. \ref{fig-shocksect}
is computed.  (f) Fidelity of the image shown in panel (e).  }
\label{fig-shock}
\end{figure*}

We plot in Figure \ref{fig-shock} the images toward the shock front
after total integrations of 10 hr for 12m$\times$50 and 40 hr for
7m$\times$12 and SD$\times$4.  Again, reconstruction is significantly
improved by adding the short baseline data of 7m$\times$12 and
SD$\times$4.

\begin{figure}[t]
  \begin{center}
    \FigureFile(80mm,50mm){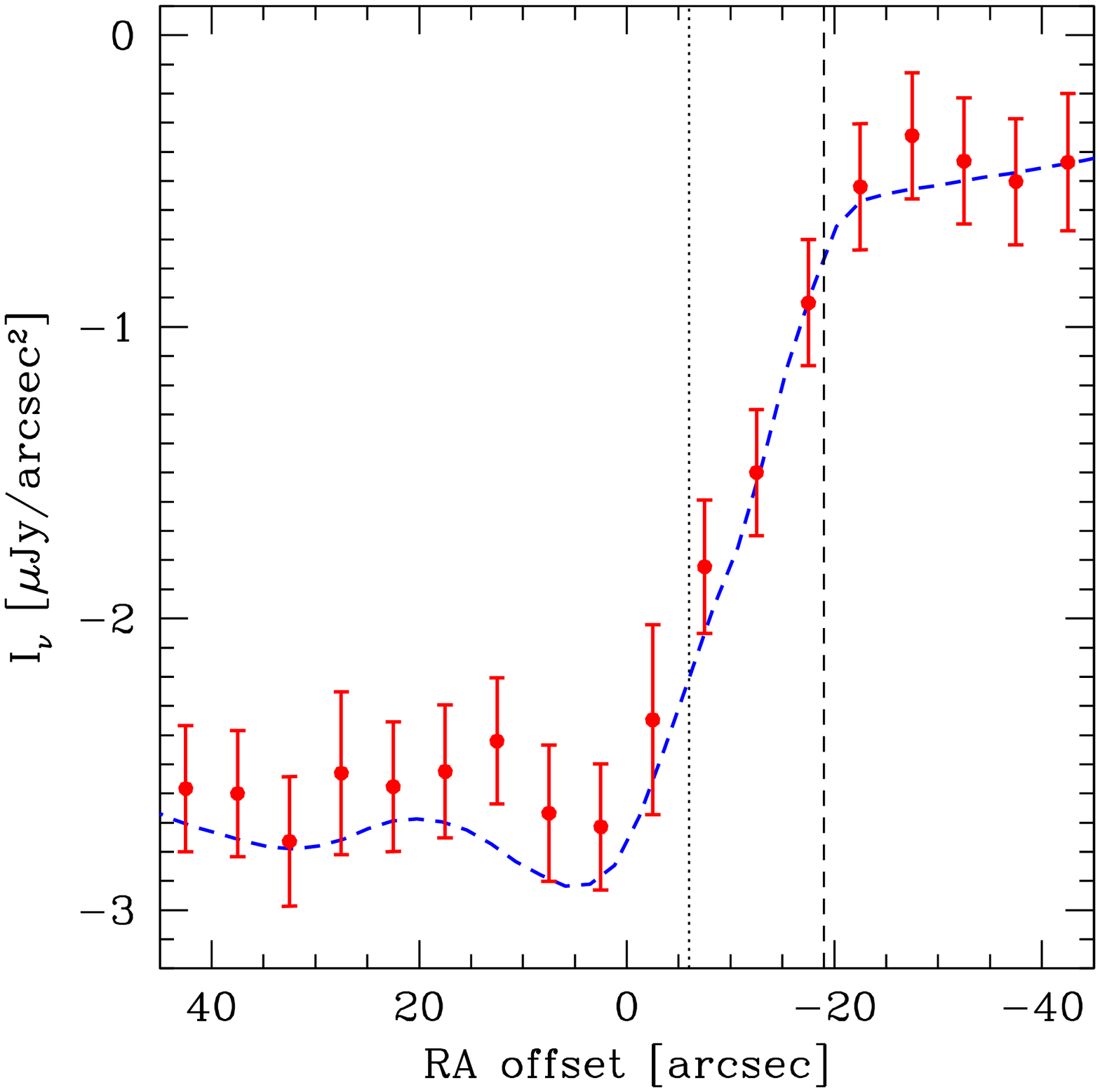}
    \FigureFile(80mm,50mm){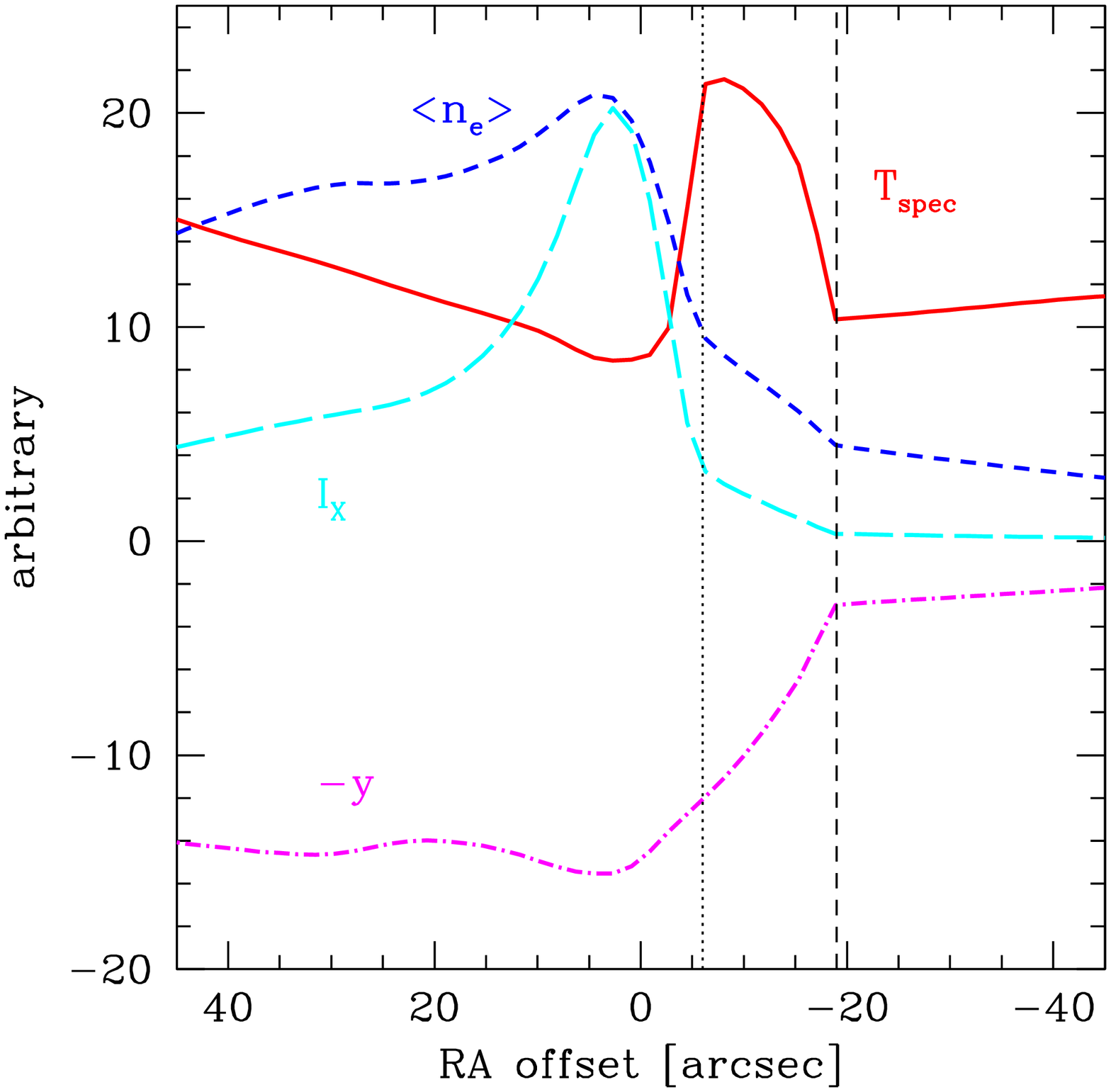}
  \end{center}
  \caption{{\it Top:} Same as Fig.~\ref{fig-bulletsect} except that the
  results of Model C shown in Fig.~\ref{fig-shock}(e) are plotted with a
  smaller bin size of $5''$. {\it Bottom:} Physical quantities with
  arbitrary units of the input model for the same region as plotted in
  the top panel. Thick lines indicate the projected temperature proposed
  by \citet{Mazzotta04} that mimics the X-ray spectroscopic temperature
  (solid), the mass weighted electron density (short dashed), the
  thermal bremsstrahlung X-ray brightness (long dashed), and the Compton
  y-parameter with a negative sign added for comparison with the SZE
  decrement (dot-dashed). } \label{fig-shocksect}
\end{figure}

Figure \ref{fig-shocksect} shows that the high angular resolution SZE
observation with ALMA is a powerful tool to resolve the shock front,
characterized by the temperature and pressure jumps. On the other hand,
the X-ray emission is weak in the low-density shock region and
rises sharply behind the contact discontinuity. The Compton y-parameter
or projected pressure is nearly constant behind the contact
discontinuity (a slight disagreement between its peak and the contact
discontinuity is due to the projection effect). The SZE and X-rays are
thus complementary in probing the detailed shock structure and the
former is particularly useful in detecting hot rarefied gas.

The width of the shock heated region is $\sim 60$ kpc corresponding to
$\sim 14''$ at $z=0.3$ as displayed in Figure \ref{fig-shocksect} or
$\sim 8''$ at $z=1$.  This physical size is consistent with the
X-ray data of this cluster presented in Figure 32 of
\citet{Markevitch07}, whereas the errors in the inferred X-ray
spectroscopic temperatures are still large.  Given the fact that the
intensity of the SZE is redshift independent and the angular diameter
distance is nearly constant at $z>1$, our results strongly point to the
good capability of ALMA in resolving shocks at such high redshifts.


\begin{figure}[t]
  \begin{center}
    \FigureFile(80mm,50mm){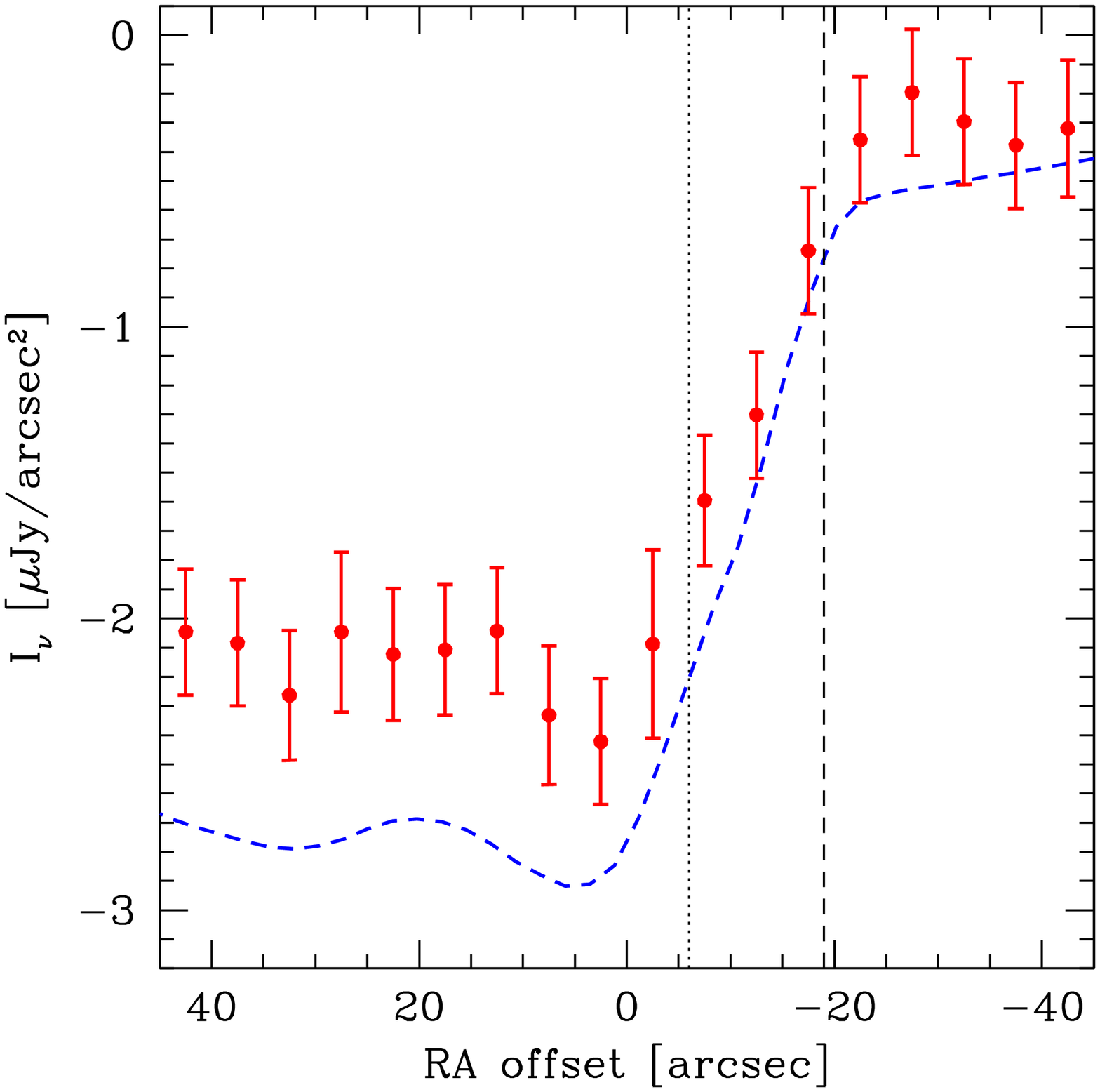}
    \FigureFile(80mm,50mm){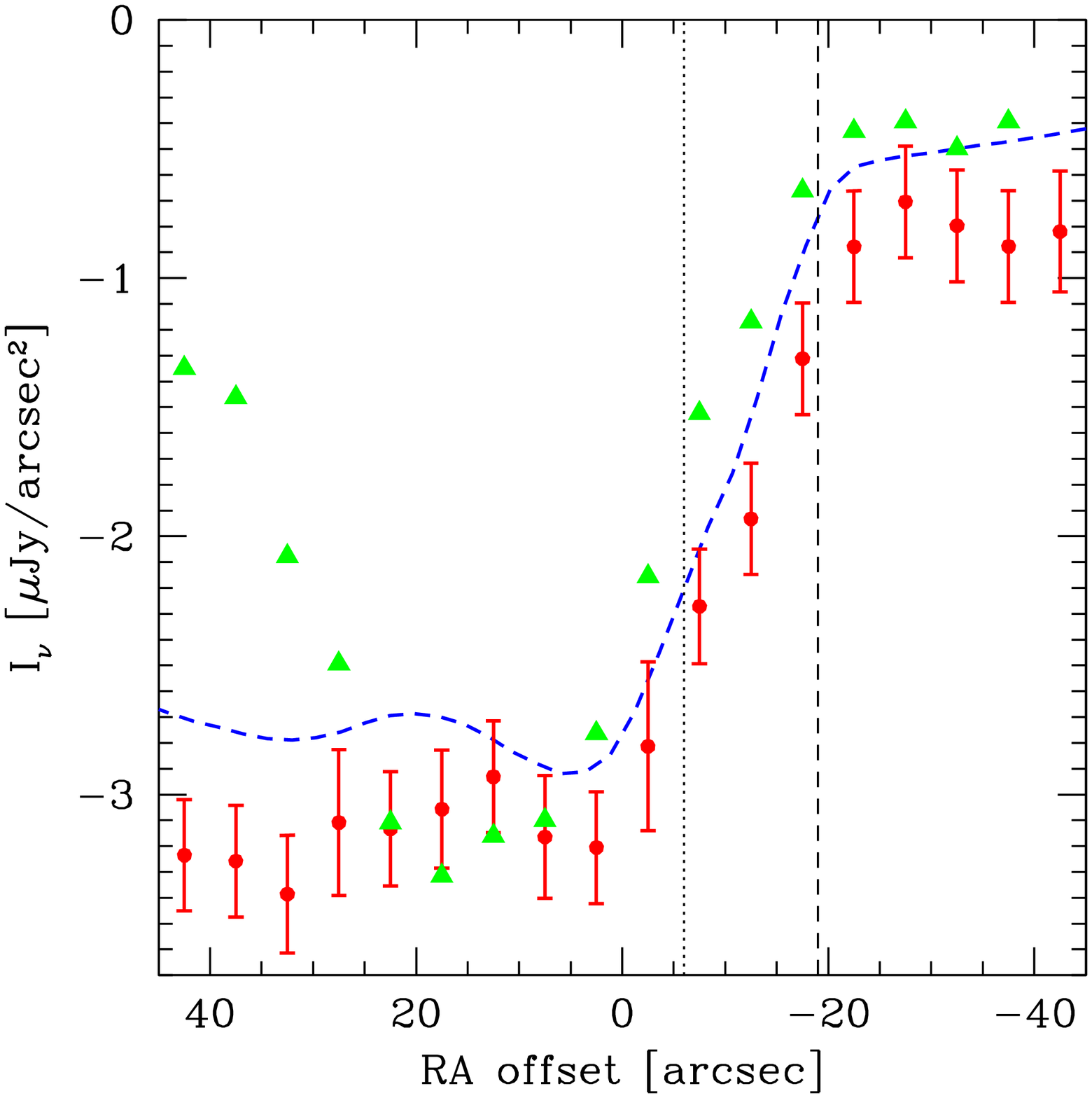}
  \end{center}
  \caption{Same as the top panel of Fig.~\ref{fig-shocksect} except for
  introducing a systematic reduction in the single-dish gain by 20\%
  (top) or additional random noise on the single-dish data that is an
  order of magnitude larger than the thermal noise (bottom). Triangles
  in the bottom panel show the results of the linear method of combining
  the single-dish data (error bars are omitted for clarity). Error bars
  do not explicitly include the single-dish noise as described in the
  text.}  \label{fig-error}
\end{figure}

\subsection{Systematic errors of the single-dish data}
\label{sec-error}

As noted in \S \ref{sec-gauss}, single-dish data play an essential
role in reconstructing the extended signals. Due to a larger beam size
and a smaller number of antennas, the single-dish data are likely to be
affected more severely by systematic errors, such as calibration errors
and atmospheric fluctuations, than the interferometric data. To check
the impacts of uniform and random systematic uncertainties, we re-run
the mock observations of Model C by i) reducing the SD$\times$4 gain by
20\% or ii) enhancing the random noise of the SD$\times$4 data by an
order of magnitude. For the latter, we also examine a linear
method, often called feathering, to combine the single-dish data instead
of non-linear joint deconvolution; a deconvolved image is produced only
for 12m$\times$50+7m$\times$12 by MEM and then combined with the
SD$\times$4 image using the MIRIAD task {\it immerge}.

Figure \ref{fig-error} shows that introducing a reduction in the
single-dish gain by $\sim 20\%$ results in a systematic decrease in the
reconstructed brightness by $\sim 20 \%$ as well. On the other hand, the
enhanced random noise primarily leads to an offset in the zero
brightness level, whereas the shape of the emission profile is nearly
unchanged. The latter bias has been known as an intrinsic problem with
MEM for low S/N data (e.g., \cite{Cornwell99}). The linear method
described above gives a better estimation of the zero level but poorer
results on the emission profile (bottom panel). This is because the
linear method directly combines the single-dish data with the
interferometer data on the $u-v$ plane and hence retains noisy pixel
values in the single-dish map. On the other hand, MEM searches for a
smooth solution that accounts for the global structure of the
single-dish data and is less sensitive to individual pixel values. In
either case, the above results suggest that the quality of the
SD$\times$4 data may limit the accuracy of the SZE observations.

\subsection{Relative integration time between the 12m array and ACA}
\label{sec-reltime}

In Models B and C, the ratio between the integration times of
12m$\times$50 and ACA (7m$\times$12 and SD$\times$4) has been taken at
its nominal value of $t_{\rm ACA}/t_{\rm 12m} = 4$ \citep{Pety01,
Morita05}. We now examine the impacts of varying this ratio by
re-running wide-field Model B simulations. The total integration time of
12m$\times$50 is fixed at 10 hr and the point sources are excluded from
the input for simplicity.

\begin{figure}[t]
  \begin{center}
    \FigureFile(80mm,50mm){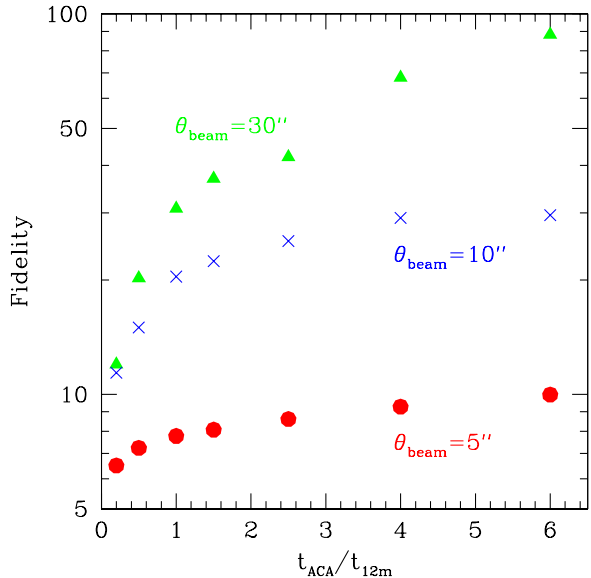}
  \end{center}
  \caption{Mean fidelity of the mock bullet cluster (Model B without
  point sources) as a function of the integration time of ACA
  (7m$\times$12 and SD$\times$4). The integration time of 12m$\times$50
  is fixed at 10 hr. The mean is taken within the central $60'' \times
  60''$ in the 12m$\times$50 + 7m$\times$12 + SD$\times$4 image without
  smoothing (circles) or after being smoothed to the effective beam FWHM
  of $10''$ (crosses) and $30''$ (triangles). }  \label{fig-reltime}
\end{figure}

Figure \ref{fig-reltime} indicates that $t_{\rm ACA}/t_{\rm 12m} \gtsim
1$ is required to achieve a $\sim 10\%$ accuracy in the image
reconstruction at the resolution of 12m$\times$50, $\theta_{\rm
beam}\sim 5''$. For smaller $t_{\rm ACA}/t_{\rm 12m}$, the thermal noise
of ACA in $\mu$Jy/arcsec$^2$ dominates over that of 12m$\times$50, as
can be readily expected from Table \ref{tab-models}.  For larger $t_{\rm
ACA}/t_{\rm 12m}$, on the other hand, the image fidelity increases more
rapidly at larger scales. The optimal value of $t_{\rm ACA}/t_{\rm 12m}$
thus depends on the spatial scale of interest, and the nominal value of
4 appears to be a reasonable choice for the case considered in the
present paper.

\section{Discussion}
\label{sec-discuss}

High resolution SZE mapping with ALMA has profound implications on
cosmology and structure formation. First of all, the shock velocity
inferred via the Rankine-Hugoniot jump condition from Chandra for 1E
0657-558 amounts to $\sim 4700$ km s$^{-1}$ \citep{Markevitch07}, which appears
to be exceptionally high within the framework of a concordance
$\Lambda$CDM model (e.g., \cite{Hayashi06,Lee10}). Our results suggest
that ALMA can significantly improve both quality and quantity of such
studies by detecting shocks up to $z>1$ and measuring directly the
pressure gap across the shock.  

The existence of very hot electrons with $kT_{\rm e} \gg 10$ keV in
clusters has also been inferred by the past SZE observation
\citep{Kitayama04} and the hard X-ray spectroscopy
\citep{Ota08,Nakazawa09,Sugawara09}. Nevertheless their spatial
distribution as well as the link to the non-thermal component are still
unclear.  ALMA will be highly complementary in probing the nature of
these relativistic electrons to hard X-ray missions such as
NuSTAR\footnote{http://www.nustar.caltech.edu/} and
ASTRO-H\footnote{http://astro-h.isas.jaxa.jp/}, whose spatial
resolutions are $45''$ and $1.7'$, respectively.

Our results can further be used to estimate the feasibility of SZE
observations in other bands.  Frequency dependence of the SZE maps
will provide a useful probe of line-of-sight gas motions via the
kinematic SZE, electron temperature via relativistic corrections, and
the existence of non-thermal electrons (e.g., \cite{Colafrancesco11,
Prokhorov11}). The SZE increment at frequencies above 600 GHz has also
been detected by Herschel toward 1E0657-558 \citep{Zemcov10}.

For given arrays and bandwidths, the thermal noise in Jy/sr is
proportional to $T_{\rm sys} t_{\rm int}^{-1/2} \nu^2$, where $T_{\rm
sys}$ is the system temperature, $t_{\rm int}$ is the integration time
and $\nu$ is the observing frequency. If one is to observe a source of
intensity $I_{\rm src}$ in Jy/sr with $N_{\rm mos}$ mosaics and perform
smoothing over $N_{\rm sm}$ beams, the signal-to-noise ratio is given by
\begin{eqnarray}
\frac{S}{N} \propto \frac{|I_{\rm src}|t_{\rm int}^{1/2}}{T_{\rm sys} \nu^2}
\frac{N_{\rm sm}^{1/2}}{N_{\rm mos}^{1/2}}. 
\end{eqnarray}
Given that both $N_{\rm mos}$ and $N_{\rm sm}$ are nearly proportional
to $\nu^2$ for fixed observing area and effective beam size after
smoothing, the integration time to reach a given signal-to-noise ratio
is
\begin{eqnarray}
t_{\rm int} \propto \frac{T_{\rm sys}^2 \nu^4}{|I_{\rm src}|^2} .
\label{eq-otherbands}
\end{eqnarray}

\begin{figure}[t]
  \begin{center}
    \FigureFile(80mm,50mm){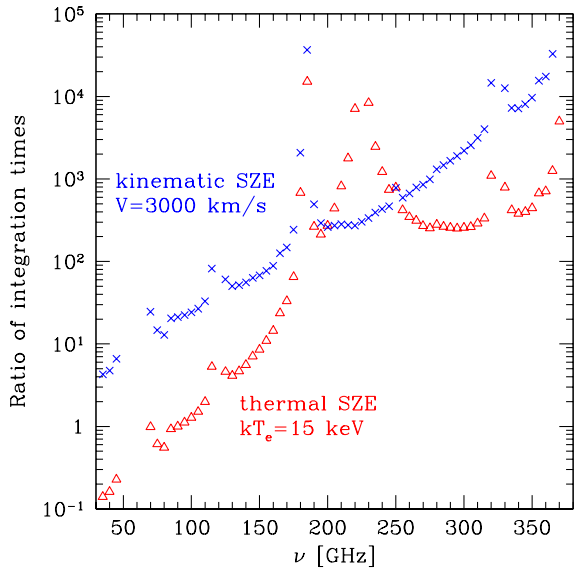}
  \end{center}
  \caption{The ratio of integration times, as a function of observing
frequency, required to reach a given signal-to-noise ratio in a fixed
effective spatial resolution over a fixed observing area.  Symbols
indicate the cases for the thermal SZE with $kT_{\rm e} = 15$ keV (triangles)
and the kinematic SZE with $V=3000$ km s$^{-1}$ (crosses).  The vertical axis
is normalized with respective to the integration time for the thermal
SZE at 90GHz. } \label{fig-otherbands}
\end{figure}

Figure \ref{fig-otherbands} shows the quantity given in equation
(\ref{eq-otherbands}) in the cases of the thermal SZE with $kT_{\rm
e}=15$ keV and the kinematic SZE with the line of sight velocity of
$V=3000$ km s$^{-1}$, including relativistic corrections
\citep{Itoh04,Nozawa05}.  The vertical axis shows the ratio with
respective to the integration time for the thermal SZE at 90GHz. We use
system temperatures at zenith $(\mbox{RA},\mbox{DEC})
=(00:00:00,-23:00:00)$ given by the ALMA sensitivity calculator.
Precipitable water vapour is taken to be 2.748 mm at $\nu<163$ GHz,
1.796 mm at $163 \leq \nu < 275$ GHz, and 1.262 mm at $275 \leq \nu <
500$ GHz. The resulting system temperatures tend to increase gradually
from $T_{\rm sys}\sim 40$ K at $\nu \sim 40$ GHz to $T_{\rm sys}\sim
180$ K at $\nu \sim 350$ GHz, with some overshoots wherever atmospheric
opacity gets high, e.g., at $\nu \sim 180$ GHz. The rise of the
integration time for the thermal SZE at $\nu \sim 220$ GHz is due to the
null of its intrinsic spectrum. It is obvious that the SZE observations,
particularly that of the kinematic SZE, become highly time-consuming at
$\nu > 100$GHz due both to an increasing $T_{\rm sys}$ and to the
$\nu^4$ factor in equation (\ref{eq-otherbands}). On the other hand,
much faster mapping is possible at lower frequencies, e.g., $\nu=40$
GHz.

Finally, we estimate the impact of unresolved ($<0.1$ mJy) point-like
sources as follows. The 150GHz source counts from recent SPT and ACT
observations are approximately $N(>S_\nu) \sim 2\times 10^{-4}
(S_{150}/10 {\rm mJy})^{-1}$ arcmin$^{-2}$ for synchrotron-dominated
sources and $N(>S_\nu) \sim 10^{-5} (S_{150}/10 {\rm ~mJy})^{-2.5}$
arcmin$^{-2}$ for dust-dominated sources at $S_{150} \sim 10$ mJy
\citep{Vieira10,Marriage11}. If we simply extrapolate these relations to
fainter fluxes at 90 GHz, the nominal confusion criterion of $1/30$
beam$^{-1}$, or equivalently $4$ arcmin$^{-2}$ for $\theta_{\rm
beam}=5''$, is reached at $S_{90} \sim 0.6 ~\mu$Jy and $10 ~\mu$Jy,
assuming the average spectral shapes of $S_\nu \propto \nu^{-0.5}$ and
$\nu^3$ for the two populations (e.g., \cite{Vieira10}), respectively.
This suggests that the source confusion is not likely to affect severely
the SZE observations considered in this paper, unless there is a
significant excess of faint sources with respect to the above
extrapolation.

\section{Conclusions}

We have performed imaging simulations of the SZE of galaxy clusters with
ALMA including ACA.  In its most compact configuration at 90GHz, ALMA
will resolve the intracluster medium with an effective angular
resolution of $\sim 5''$, corresponding to $\sim 20$ kpc at $z=0.3$ and
$\sim 40$ kpc at $z=1$. Such observations will be particularly useful in
detecting shock fronts and/or relativistic (either thermal with $kT_{\rm e}
\gg 10$ keV or non-thermal) electrons produced during violent cluster
mergers at high redshifts, that are hard to resolve by current and
near-future X-ray detectors.

Our results imply that ACA plays an essential role in reconstructing the
global structures of the SZE and its capability may be limited by the
accuracy of single-dish data.  Expected sensitivity of the 12m array
based on the thermal noise is not valid for the SZE mapping unless
accompanied by an ACA observation of at least equal duration. An optimal
ratio of integration times between the two observations depends on the
spatial scale of interest and the nominal value of 4 appears to be a
reasonable choice in the present case. The SZE observations at $\nu >
100$GHz will become excessively time-consuming owing to the narrower
beam size and the larger system temperature. On the other hand,
significant improvement of the observing efficiency is expected once
Band 1 is implemented in the future.

In addition to the SZE, ALMA will detect radio/IR sources in the
field-of-view. The spectral energy distributions of such sources in the
ALMA bands are still highly uncertain and will be crucial both in
understanding the nature of the sources and in quantifying their
contamination to the SZE survey data (e.g., \cite{Lin09}).

\bigskip

We thank Koh-Ichiro Morita, Yasutaka Kurono, Tetsuo Hasegawa and Ray
Furuya for helpful discussions, and Andy Biggs for his correspondence on
the ALMA sensitivity calculator. We also thank the anonymous referee for
useful comments. This work is supported in part by the Grants-in-Aid for
Scientific Research by the Japan Society for the Promotion of Science
(21740139, 20340041), the Global Scholars Program of Princeton
University, the grant from the National Science Council of Taiwan
(99-2112-M-001-009-MY3), and the Korea Research Council of Fundamental
Science and Technology research fellowship for young scientist.
Numerical computations were in part carried out on XT4 at the Center for
Computational Astrophysics, CfCA, of the National Astronomical
Observatory of Japan.

\section*{Note Added in Proof (June 22, 2012)}

After this paper was accepted for publication, updates to the ALMA
sensitivity calculator released on May 31, 2012 resulted in an increase
in the system temperature at 90GHz by $10 \%$ and an inclusion of
quantization efficiency of $0.96$, compared to those listed in Table
\ref{tab-telesc}.  These modifications would simply raise the thermal
noise amplitude by $14\%$ and the results of the present paper remain
essentially unchanged once the quoted integration times are scaled
upward by $30\%$.

\appendix
\section{Effective Integration Time in Mosaicing Observations}
\label{sec-append}

In mosaicing observations, effective integration time toward a given
sky position $\vec{\theta}$ is expressed as
\begin{equation}
t_{\rm eff}(\vec{\theta}) = \epsilon_{\rm m}^2(\vec{\theta}) \frac{t_{\rm int}}{N_{\rm mos}},
\label{eq-teff}
\end{equation}
where $t_{\rm int}$ is the total integration time over the entire
mapping area, $N_{\rm mos}$ is the number of mosaics, and $\epsilon_{\rm
m}(\vec{\theta})$ is the ``mosaicing overlap factor'' introduced in
\citet{Holdaway95} and \citet{Morita05}. The last quantity accounts for
the overlap of mosaics that cover the same position $\vec{\theta}$ and
is given by
\begin{equation}
\epsilon_{\rm m}^2(\vec{\theta}) = \sum_{p=1}^{N_{\rm mos}} g^2(\vec{\theta}-\vec{\theta}_p) ,
\label{eq-xi}
\end{equation}
where $g$ is the primary beam gain and $\vec{\theta}_p$ is the pointing
center of the $p$-th mosaic. In what follows, we derive the values of
$\langle \epsilon_{\rm m}^2 \rangle $ and $\langle t_{\rm eff} \rangle$
analytically.

First, we consider primary beam gain functions of the form
\begin{eqnarray}
g(\theta) = \left\{
\begin{array}{ll}
\exp\left(-\frac{\theta^2}{2\sigma_{\rm G}^2} \right)
& \mbox{Gaussian,} \\
\left[\frac{2 J_1(\pi D \theta/\lambda)}{\pi D \theta/\lambda} \right]^2
& \mbox{Airy disk,}
\end{array}
\right.
\end{eqnarray}
where $J_1$ is the Bessel function of the the first kind of order unity,
$D$ is the diameter of the telescope, and $\lambda$ is the observing
wavelength. Integrating $g^2$ over the sky gives the effective area of
integration per pointing as
\begin{eqnarray}
A_{\rm gain} = 
\left\{ \begin{array}{ll}
\pi \sigma_{\rm G}^2 \simeq 0.567 ~\theta_{\rm pb}^2 & \mbox{Gaussian,} \\
\frac{2^5 \lambda^2 I_{\rm A}}{\pi D^2} \simeq 0.554 ~\theta_{\rm pb}^2 
& \mbox{Airy disk,}
\end{array}
\right.
\label{eq-again}
\end{eqnarray}
where 
\begin{equation}
I_{\rm A} \equiv \int_0^\infty \frac{J_1(x)^4}{x^3} dx \simeq 0.0575, 
\end{equation}
and $\theta_{\rm pb}$ is the primary beam FWHM corresponding to $\sqrt{8
\ln 2} \sigma_{\rm G}$ for Gaussian and $1.028 \lambda/D$ for the Airy
disk.  For each $p$ in equation (\ref{eq-xi}), $g^2 \sim 1$ within
$A_{\rm gain}$ centered at $\vec{\theta}_p$, whereas $g^2 \sim 0$
otherwise.

Second, we suppose that mosaicing centers are placed on a regular grid,
either square or triangular, with constant spacing $d$.  The physical
area around each grid point enclosed by perpendicular bisectors with the
neighboring points is
\begin{eqnarray}
A_{\rm grid} = \left\{ \begin{array}{ll}
d^2 & \mbox{square grid,} \\
\frac{\sqrt{3}}{2}d^2 & \mbox{triangular grid,}
\end{array}
\right.
\label{eq-agrid}
\end{eqnarray}
where the shape of the enclosed area is a square and a hexagon,
respectively. 

Finally, we average equation (\ref{eq-xi}) around an arbitrary grid point
$\vec{\theta}_q$ that lies sufficiently far ($\gg \sqrt{A_{\rm
gain}/\pi}$) from the map edge, to obtain
\begin{eqnarray}
\langle \epsilon_{\rm m}^2 \rangle &=&
\frac{\sum_{p=1}^{N_{\rm mos}} \int_q g^2(\vec{\theta}-\vec{\theta}_p) d^2\theta}{\int_{q} d^2\theta} \nonumber \\
&=& \frac{\sum_{p=1}^{N_{\rm mos}} \int_p g^2(\vec{\theta}-\vec{\theta}_q) d^2\theta}{\int_q d^2\theta} 
= \frac{A_{\rm gain}}{A_{\rm grid}}, 
\label{eq-xibar1}
\end{eqnarray}
where $\int_i$ denotes an integral around the $i$-th grid point over the
area $A_{\rm grid}$.  The last result in equation (\ref{eq-xibar1}) does
not depend on $q$ and remains unchanged even if the average is taken
over multiple grids. Substituting equations (\ref{eq-again}) and
(\ref{eq-agrid}) into (\ref{eq-xibar1}) yields
\begin{eqnarray}
\langle \epsilon_{\rm m}^2 \rangle &=&  \langle \epsilon_{\rm m}^2
\rangle_{\rm Ny} 
\left(\frac{d}{0.5 \theta_{\rm pb}}\right)^{-2}, 
\label{eq-xibar2}
\end{eqnarray}
where the representative values in the case of Nyquist spacing are
\begin{eqnarray}
\langle \epsilon_{\rm m}^2
\rangle_{\rm Ny}  &\simeq & \left\{ \begin{array}{ll}
2.27 & \mbox{Gaussian, square grid,}\\
2.62 & \mbox{Gaussian, triangular grid,}\\
2.22 & \mbox{Airy disk, square grid,}\\
2.56 & \mbox{Airy disk, triangular grid.}
\end{array}
\right.
\label{eq-xibar3}
\end{eqnarray}
Equations (\ref{eq-teff}) and (\ref{eq-xibar1}) give 
\begin{equation}
\frac{\langle t_{\rm eff} \rangle}{t_{\rm int}} = 
\frac{\langle \epsilon_{\rm m}^2 \rangle}{N_{\rm mos}} 
\sim \frac{A_{\rm gain}}{A_{\rm map}},
\label{eq-teff2}
\end{equation}
where $N_{\rm mos} A_{\rm grid}$ roughly equals the mapping area $A_{\rm
map}$.  Equation (\ref{eq-teff2}) indicates that the {\it mean}
integration time at each sky position within fixed $A_{\rm map}$ {\it
does not} (apart from the time spent at the map edge) depend on $d$ or
the grid orientation; $\langle \epsilon_{\rm m}^2 \rangle$ and $N_{\rm
mos}$ are both proportional to $d^{-2}$.  On the other hand, its {\it
dispersion does} as shown below.

\begin{figure}[t]
  \begin{center}
    \FigureFile(80mm,50mm){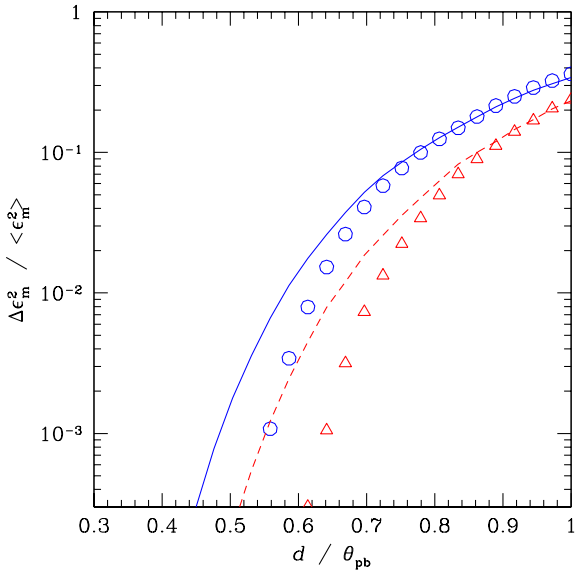}
  \end{center}
  \caption{Standard deviation of the mosaicing overlap factor $\Delta
  \epsilon_{\rm m}^2$ as a function of grid spacing $d$ for a Gaussian
  primary beam on square (solid line) and triangular grids (dashed line)
  or an Airy-disk-shaped primary beam on square (circles) and triangular
  grids (triangles). The vertical and horizontal axes are normalized by
  the mean $\langle \epsilon_{\rm m}^2 \rangle$ and the primary beam
  FWHM $\theta_{\rm pb}$, respectively.  } \label{fig-mosfac}
\end{figure}

We further compute numerically the mean and the standard deviation of
$\epsilon_{\rm m}^2$ by creating square and triangular grids over the
sky area with $5 \theta_{\rm pb} \times 5 \theta_{\rm pb}$ and
extracting 10000 random positions of $\vec{\theta}$ from the central $3
\theta_{\rm pb} \times 3 \theta_{\rm pb}$. Figure \ref{fig-mosfac}
illustrates that $\Delta \epsilon_{\rm m}^2 / \langle \epsilon_{\rm m}^2
\rangle $, or equivalently $\Delta t_{\rm eff}/\langle t_{\rm eff}
\rangle$, drops sharply to $\ll 1\%$ at $d < 0.5 \theta_{\rm pb}$ in all
the cases considered here. We have checked that $\langle \epsilon_{\rm
m}^2 \rangle$ computed numerically agrees with equations
(\ref{eq-xibar2}) and (\ref{eq-xibar3}) to better than $1\%$ and $4\%$
at $d<0.7 \theta_{\rm pb}$ and $d< \theta_{\rm pb}$, respectively. The
agreement improves further once the larger computational area is
adopted.

The above results imply that, as far as the Nyquist condition is
satisfied, square and triangular grids lead to practically similar and
uniform effective integration time apart from the map edge.  The
difference due to assumed shapes of the primary beam, Gaussian or an
Airy disk, is not significant either.



\begin{thebibliography}{}

\bibitem[Akahori \& Yoshikawa(2010)]{Akahori10}
Akahori, T., \&  Yoshikawa, K. 2010, PASJ, 62, 335

\bibitem[Akahori \& Yoshikawa(2012)]{Akahori12}
Akahori, T., \&  Yoshikawa, K. 2012, PASJ, 64, 12 

\bibitem[AMI Collaboration (2006)]{AMI06}
AMI Collaboration: Barker, R., et al. 2006, MNRAS, 369, L1 

\bibitem[Andreani et al.(1999)]{Andreani99}
Andreani, P., et al. 1999, ApJ, 513, 23

\bibitem[Birkinshaw(1999)]{Birkinshaw99} 
Birkinshaw, M.\ 1999, Phys. Rep. 310, 97

\bibitem[Briggs(1995)]{Briggs95}
Briggs, D. S. 1995, PhD theses, New Mexico Tech. 

\bibitem[Carlstrom, Holder \& Reese(2002)]{Carlstrom02} 
Carlstrom, J. E., Holder, G. P., \& Reese, E. D., 2002, ARA\&A, 40, 643

\bibitem[Carlstrom, Joy \& Grego(1996)]{Carlstrom96} 
Carlstrom, J. E., Joy, M., Grego, L. 1996, 456, L75

\bibitem[Colafrancesco, Marchegiani  \& Buonanno (2011)]{Colafrancesco11}
Colafrancesco, S., Marchegiani, P., \& Buonanno, R. 2011, A\&A, 527, L1


\bibitem[Cornwell, Braun \& Briggs(1999)]{Cornwell99} 
Cornwell, T., Braun, R., \& Briggs, D. S. 1999, 
in ASP Conf. Ser. 180, Synthesis Imaging in Radio Astronomy II,
ed. G. B. Taylor, C. L. Carilli, \& R. A. Perley (San
Francisco: ASP), 151

\bibitem[Muchovej et al.(2007)]{Muchovej07}	
Muchovej, S., et al. 2007, ApJ, 663, 708

\bibitem[Halverson et al.(2009)]{Halverson09} 
Halverson, N. W., et al. 2009, ApJ, 701, 42

\bibitem[Hayashi \& White(2006)]{Hayashi06} 
Hayashi, E., \& White, S. D. M. 2006, MNRAS, 370, L380

\bibitem[Helfer et al.(2002)]{Helfer02} 
Helfer, T. T., Vogel, S. N., Lugten, J. B., \& Teuben, P. J. 2002, PASP, 
114, 793

\bibitem[Holdaway \& Rupen(1995)]{Holdaway95}
Holdaway, M. A., \& Rupen M. P. 1995, ALMA Memo 128


\bibitem[Iguchi et al.(2009)]{Iguchi09}
Iguchi, S., et al.  2009, PASJ, 61, 1 

\bibitem[Itoh \& Nozawa(2004)]{Itoh04} 
Itoh, N., \&  Nozawa, S. 2004, A\&A, 417, 827

\bibitem[Jones et al.(1993)]{Jones93} 
Jones, M., et al. 1993, Nature, 365, 320 

\bibitem[Kitayama et al.(2004)]{Kitayama04} 
Kitayama, T., Komatsu, E., Ota, N., Kuwabara, T., Suto, Y.,
Yoshikawa, K., Hattori, M., \& Matsuo, H. 2004, PASJ, 56, 17 

\bibitem[Komatsu et al.(1999)]{Komatsu99} 
Komatsu, E., Kitayama, T., Suto, Y., Hattori, M., Kawabe, R.,
Matsuo, H., Schindler, S., \& Yoshikawa, K.\ 1999, ApJ, 516, L1

\bibitem[Komatsu et al.(2001)]{Komatsu01} 
 Komatsu, E., et al. 2001, PASJ, 53, 57

\bibitem[Komatsu et al.(2011)]{Komatsu11}
 Komatsu, E., et al. 2011, ApJS, 192, 18

\bibitem[Korngut et al.(2011)]{Korngut11}
Korngut, P. M., et al. 2011, ApJ, 734, 10

\bibitem[Kurono, Morita \& Kamazaki(2009)]{Kurono09}
Kurono, Y., Morita, K., \& Kamazaki, T. 2009, PASJ, 61, 873

\bibitem[Lee \& Komatsu(2010)]{Lee10}
Lee, J., Komatsu, E. 2010, ApJ, 718, 60 

\bibitem[Liang et al.(2000)]{Liang00}
Liang, H., Hunstead, R. W., Birkinshaw, M., \& Andreani, P. 2000, ApJ, 544, 686

\bibitem[Lin et al.(2009)]{Lin09}
Lin, Y.-T., Partridge, B., Pober, J. C., Bouchefry, K. E., 
Burke, S., Klein, J. N., Coish, J. W., Huffenberger, K. M., 2009, 
\apj, 694, 992 


\bibitem[Malu et al.(2010)]{Malu10}
Malu, S. S., Subrahmanyan, R., Wieringa, M., 
\& Narasimha, D. 2010, arXiv:1005.1394

\bibitem[Markevitch et al.(2002)]{Markevitch02}
Markevitch, M., Gonzalez, A. H., David, L., Vikhlinin, A., 
Murray, S., Forman, W., Jones, C., Tucker, W. 2002, ApJ, 567, L27

\bibitem[Markevitch \& Vikhlinin(2007)]{Markevitch07}
Markevitch, M., \& Vikhlinin, A., 2007, 
Physics Reports, 443, 1

\bibitem[Marriage et al.(2011)]{Marriage11}
Marriage, T. A., et al. 2011 ApJ, 731, 100

\bibitem[Massardi et al.(2010)]{Massardi10}
Massardi, M., Ekers, R. D., Ellis, S. C., Maughan, B. 2010, ApJ, 718, L23

\bibitem[Mason et al.(2010)]{Mason10}
Mason, B. S., et al. 2010, ApJ, 716, 739 

\bibitem[Mazzotta et al.(2004)]{Mazzotta04}
Mazzotta, P., Rasia, E., Moscardini, L., \& Tormen, G. 2004, MNRAS, 354, 10
 
\bibitem[Morita \& Holdaway(2005)]{Morita05}
Morita, K., \& Holdaway, M. A. 2005, ALMA Memo  538

\bibitem[Nakazawa et al.(2009)]{Nakazawa09}
Nakazawa, K., et al. 2009, PASJ, 61, 339

\bibitem[Narayan \& Nityananda(1986)]{Narayan86}
Narayan, R., \&  Nityananda, R. 1986, ARA\&A 24, 127

\bibitem[Nozawa, Itoh \& Kohyama(2005)]{Nozawa05} 
Nozawa, S., Itoh, N., Kohyama, Y. 2005, A\&A, 440, 39

\bibitem[Ota et al.(2008)]{Ota08}
Ota, N., et al. 2008, A\&A, 491, 363 

\bibitem[Pety, Gueth \& Guilloteau(2001)]{Pety01}
Pety, J. , Gueth F., \& Guilloteau, S. 2001, ALMA Memo  398

\bibitem[Pfrommer, Ensslin \& Sarazin(2005)]{Pfrommer05}
Pfrommer, C., Ensslin, T. A., \& Sarazin, C. L. 2005, A\&A, 430, 799

\bibitem[Plagge et al.(2010)]{Plagge10}
Plagge, T., et al. 2010, ApJ, 716, 1118

\bibitem[Pointecouteau et al.(2001)]{Pointecouteau01}
Pointecouteau, E., Giard, M., Benoit, A., D\'esert, F. X., 
Bernard J.P., Coron, N., \& Lamarre, J. M. \ 2001, ApJ, 552, 42

\bibitem[Prokhorov et al.(2011)]{Prokhorov11}
Prokhorov, D. A., Colafrancesco, S., Akahori, T., Million, E. T., Nagataki, S., Yoshikawa, K. 2011, MNRAS, 416, 302


\bibitem[Rephaeli(1995)]{Rephaeli95} 
Rephaeli, Y.\ 1995, ARA\&A, 33, 541

\bibitem[Sugawara, Takizawa \& Nakazawa(2009)]{Sugawara09}
Sugawara, C., Takizawa, M., \& Nakazawa, K. 2009, PASJ, 61, 1293


\bibitem[Sault(1990)]{Sault90}
Sault, R. J. 1990, ApJ, 354, L61

\bibitem[Sault, Staveley-Smith \& Brouw(1996)]{Sault96}
Sault, R. J., Staveley-Smith, L., \& Brouw, W. N. 1996, A\&AS, 120, 375

\bibitem[Sault, Teuben \& Wright(1995)]{Sault95}
Sault R. J., Teuben P. J., \& Wright M. C. H. 1995,  
In Astronomical Data Analysis Software and Systems IV, 
eds R. Shaw, H.E. Payne, J.J.E. Hayes, 
ASP Conference Series, 77, 433


\bibitem[Steer, Dewdney \& Ito(1984)]{Steer84}
Steer, D. G., Dewdney, P. E., \& Ito, M. R. 1984, 
A\&A, 137, 159

\bibitem[Stanimirovic et al.(1999)]{Stanimirovic99}
Stanimirovic, S., Staveley-Smith, L., Dickey, J. M., 
Sault, R. J., Snowden, S. L. 1999, MNRAS, 302, 417 

\bibitem[Sunyaev \& Zel'dovich(1972)]{Sunyaev72} 
Sunyaev, R. A., \& Zel'dovich, Ya. B.\ 1972, 
Comments Astrophys. Space
Phys.,  4, 173

\bibitem[Takakuwa et al.(2003)]{Takakuwa03}
Takakuwa, S., Kamazaki, T., Saito, M., \& Hirano, N. 2003, ApJ, 584, 818


\bibitem[Takakuwa et al.(2008)]{Takakuwa08}
Takakuwa, S., Iono, D., Vila-Vilaro, B., Sekiguchi, T., 
\& Kawabe, R. 2008, Ap\&SS, 313, 169

\bibitem[Takizawa(2005)]{Takizawa05} 
Takizawa, M. 2005, ApJ, 629, 791 

\bibitem[Vieira et al.(2010)]{Vieira10}
Vieira, J. D., et al. 2010, ApJ, 719, 763

\bibitem[Vogel et al.(1984)]{Vogel84}
Vogel, S. N., Wright, M. C. H., Plambeck, R. L., \& Welch, W. J. 1984,
ApJ, 283, 655

\bibitem[Yen et al.(2011)]{Yen11}
Yen, H-W., Takakuwa, S., Ohashi, N. 2011, ApJ, 742, 57

\bibitem[Wilson et al.(2008)]{Wilson08}
Wilson, G. W., et al. 2008, MNRAS, 390, 1061 

\bibitem[Wu et al.(2009)]{Wu09}
Wu, J-H. P., et al. 2009, ApJ, 694, 1619  

\bibitem[Zemcov et al.(2010)]{Zemcov10}
Zemcov, M., et al. 2010, A\&A, 518, L16

\end{thebibliography}
\end{document}